\documentclass[twocolumn, pra, superscriptaddress]{revtex4-2}

\usepackage{graphicx}
\usepackage{dcolumn}
\usepackage{bm}

\usepackage{amsfonts}
\usepackage{amssymb}
\usepackage{amsmath}
\usepackage{calc}
\usepackage{graphicx}
\usepackage{bm}
\usepackage{verbatim}
\usepackage{float}

\def\be{ \begin{equation} }
\def\ee{ \end{equation} }
\def\bea{ \begin{align} }
\def\eea{ \end{align} }
\def\bse{ \begin{subequations} }
\def\ese{ \end{subequations} }

\def\ket#1{\vert #1 \rangle}
\def\bra#1{\langle #1 \vert}

\begin{document}


\title{Dynamics of Fock states in traveling pulses}


\author{Vasil Z. Vasilev}
\affiliation{Center for Quantum Technologies, Department of Physics, Sofia University, 5 James Bourchier Boulevard, 1164 Sofia, Bulgaria}
\author{Benedetto Militello}
\affiliation{Universita degli Studi di Palermo, Dipartimento di Fisica e Chimica - Emilio Segre, Via Archirafi 36, 90123 Palermo, Italy}
\affiliation{INFN Sezione di Catania, Via Santa Sofia 64, 95123 Catania, Italy}
\author{Anna Napoli}
\affiliation{Universita degli Studi di Palermo, Dipartimento di Fisica e Chimica - Emilio Segre, Via Archirafi 36, 90123 Palermo, Italy}
\affiliation{INFN Sezione di Catania, Via Santa Sofia 64, 95123 Catania, Italy}
\author{Nikolay V. Vitanov}
\affiliation{Center for Quantum Technologies, Department of Physics, Sofia University, 5 James Bourchier Boulevard, 1164 Sofia, Bulgaria}


\date{\today}

\begin{abstract}
The quantum state of a traveling light pulse can qualitatively change the excitation dynamics of a two-level atom. We investigate this dependence for coherent and Fock input states using a virtual-cavity formulation with matched input and output temporal profiles. Analytical treatment of the master equation and numerical simulations show that coherent pulses produce damped Rabi oscillations, whereas Fock pulses can exhibit a partial recovery of the oscillation amplitude in the trailing part of the pulse. We examine how coupling to an auxiliary temporal mode and radiative loss shape this recovery. We also establish a connection with stimulated Raman adiabatic passage through an exact many-photon dark state and derive the associated adiabaticity condition. The observed recovery occurs away from ideal dark-state following. These results distinguish transient recovery in traveling-pulse scattering from conventional cavity collapse and revival, and clarify the scope of an adiabatic-transfer interpretation.
\end{abstract}


\maketitle

\section{Introduction \label{Sec:intro}}
The interaction of an electromagnetic field with an atom is a standard subject in quantum optics textbooks. The simplest description uses a semiclassical model in which the light pulse is treated classically and the atom is treated quantum mechanically as a two-level system \cite{gerryknight}.
The Jaynes-Cummings model successfully describes the situation where an atom passes through a cavity prepared in a given state. In quantum control, however, the atom often remains stationary while a light pulse, generated by a laser or another source, passes through it. This situation is fundamentally different from the semiclassical one, in that the field now has a multimode character (an infinite number of frequency modes is needed to describe a traveling wave packet). The outgoing field is also multimode, with mode populations that may differ from those of the incident pulse.

The quantum state of the incident pulse introduces a further distinction. Pulses with the same temporal profile and mean photon number can have different photon-number fluctuations and produce different atom-field correlations. Coherent and Fock states provide a natural comparison: the former has Poissonian photon statistics, whereas the latter has a definite photon number. Their comparison therefore reveals features of the atomic response that cannot be inferred from the mean incident intensity alone. Establishing how these features are affected by scattering and spontaneous emission is important for controlling atomic excitation with quantized radiation. Atomic excitation by propagating Fock and coherent pulses has been studied for single-photon input in Ref.~\cite{Wang2011excitation} and for multiphoton input in two spatial modes in Ref.~\cite{Wang2012stateDependent}. In the complementary setting of an initially excited atom, exact pulsed stimulated emission and the large-photon-number Fock-drive regime were studied in Ref.~\cite{Fischer2018stimulated}.

Solving this problem is computationally demanding because of the infinite-dimensional field Hilbert space. However, Kiilerich and M\o lmer proposed a simpler approach in Ref.~\cite{molmer_2019}: if we associate the input and output pulses with single temporal modes $u(t)$ and $v(t)$, respectively, the traveling-pulse problem can be formulated in the tensor-product Hilbert space of a virtual input cavity, the system of interest, and a virtual output cavity, using Jaynes-Cummings-type couplings.

The collapse and revival of atomic Rabi oscillations in the Jaynes-Cummings model were established in Ref.~\cite{Eberly1980collapse} and analyzed in detail in Ref.~\cite{Narozhny1981coherence}. Of particular interest is the partial recovery of atomic oscillations in the trailing part of a Fock-state pulse. Its origin is not evident from the pulse envelope alone. A physical explanation must account for the correlations generated during scattering and distinguish this transient recovery from conventional cavity collapse and revival.

In this work, we study how the input-cavity state affects the excitation dynamics of the atom. Section~\ref{Sec:background} presents the underlying models. In section \ref{Sec:derivation}, we derive the equations for coherent and Fock input states. In section \ref{Sec:results}, we present QuTiP simulations. Section \ref{Sec:stirap} examines the analogy between a traveling pulse and STIRAP in a three-level system. We examine the adiabaticity conditions and their relevance to the late-pulse recovery of the atomic oscillations. Finally, in section \ref{Sec:conclusions}, we give concluding remarks and suggest further research directions.

\section{Background \label{Sec:background}}
In the semiclassical model (often called the Rabi model), the total Hamiltonian has the form
\be
\label{rabi-hamiltonian}
\hat{H}(t) = \hat{H}_S(t) + i [u(t) \hat{c} - u(t)^* \hat{c}^\dag],
\ee
where $\hat{H}_S$ is the free-system Hamiltonian and the rest is the interaction term. 
The light is represented by the pulse shape factor $u(t)$. The main advantage of this model is its simplicity, which allows one to find exact analytical solutions for different types of interaction (different forms of $u(t)$ and different detunings), but it works only under the assumption of a purely coherent input field. It cannot explain phenomena related to the quantum nature of light, such as the collapse-and-revival phenomenon.

A more complete description of light-atom interaction is obtained by treating the electromagnetic field quantum mechanically with its raising and lowering operators $\hat{a}^\dag$, $\hat{a}$. In the Jaynes-Cummings model, the light is assumed to occupy only a single mode, as in a cavity where the frequencies of all other modes are far from the mode of interest (by much more than the detuning from the atomic transition of the cavity mode). A typical Hamiltonian in the rotating-wave approximation is
\be
\label{jc-hamiltonian}
\hat{H}(t) = \hat{H}_S(t) + i [g^*(t) \hat{a}^\dag \hat{c} - g(t) \hat{a}\hat{c}^\dag],
\ee
where the system Hamiltonian $\hat{H}_S$ now includes the free atomic and cavity terms, while the interaction term describes photon absorption or emission when the atom is excited or deexcited. In the lossless model, the photon-number components of a coherent cavity field oscillate at different Rabi frequencies; their dephasing and subsequent rephasing produce collapse and revival, as described in Refs.~\cite{Eberly1980collapse,Narozhny1981coherence}.

 Interaction with the environment can be included through Lindblad operators, whose simplest forms for a zero-temperature bath are
\be
\label{jc-hamiltonian}
\hat{L}_c = \sqrt{\gamma}\hat{c},\quad
\hat{L}_a = \sqrt{\delta}\hat{a}.
\ee

For the virtual-cavity description introduced above, the Hamiltonian has the form
\be
\label{travel-hamiltonian}
\begin{split}
\hat{H}(t) & = \hat{H}_S(t) + \frac{i}{2} [\sqrt{\gamma}g^*_u(t) \hat{a}^\dag_u \hat{c} + \sqrt{\gamma^*}g_v(t) \hat{c}^\dag\hat{a}_v \\
& + g^*_u(t)g_v(t)\hat{a}^\dag_u\hat{a}_v - h.c.].
\end{split}
\ee
Here $\hat{a}_u$ and $\hat{a}_v$ are the lowering operators of the input and output cavities, respectively, and $\hat{c}$ is the lowering operator of the system. The couplings $g_u(t)$ and $g_v(t)$ are chosen so that the input cavity can emit a pulse with profile $u(t)$ and the output cavity can absorb a pulse with profile $v(t)$. They can be written as follows:
\begin{subequations}
\begin{align}
\label{g_u_t_form}
g_u(t) &= \frac{u^*(t)}{\sqrt{1-\int_{0}^{t} dt' |u(t')|^2}},
\\
\label{g_v_t_form}
g_v(t) &= -\frac{v^*(t)}{\sqrt{\int_{0}^{t} dt' |v(t')|^2}}.
\end{align}
\end{subequations}
It is important to note that $u(t)$ and $v(t)$ describe normalized single-photon temporal profiles, i.e., their squared moduli integrate to 1. The coefficient $\gamma$ determines both the coupling of the input pulse to the system and the rate at which the system loses excitation into the output. The optimal value of $\gamma$ is determined by the temporal length of the pulse $\tau$, $\gamma \sim 1/\tau$.

The Hamiltonian in Eq.~\eqref{travel-hamiltonian} is complemented by a single Lindblad jump operator that is responsible for deexcitation of the three components: the input cavity, the system, and the output cavity,
\be
\label{travel-lidblad}
\hat{L}_0(t) = \sqrt{\gamma}\hat{c} + g_u(t)\hat{a}_u + g_v(t)\hat{a}_v.
\ee
Additional $m$ dissipation channels can be included through additional Lindblad operators, $\hat{L}_j(t), j > 0$. The resulting master equation is
\be
\label{travel-motion}
\frac{d\rho}{dt} = -i[\hat{H}, \rho] + \sum_{j=0}^m(\hat{L}_j\rho\hat{L}_j^\dag - \frac{1}{2}\{\hat{L}_j^\dag\hat{L}_j, \rho\}).
\ee
 Substituting $\hat{H}$ and $\hat{L}_0$ into this equation shows that population is transferred only from left to right \cite{molmer_2020}, which is the essence of the input-output framework.

The above description can be generalized to several input and output pulse modes, as described in Ref.~\cite{molmer_2020}, but here we consider only the single-mode case. An alternative formulation based on a hierarchy of master equations for an arbitrary quantum system driven by an $N$-photon wave packet was developed in Ref.~\cite{Baragiola2012nphoton}.

 The most appropriate output-pulse profile depends on the eigenmode decomposition of the $\hat{L_0}$ autocorrelation function, as described in Ref.~\cite{molmer_2020}. Only the modes with the largest eigenvalues need be retained, but if the system is a two-level atom, the most probable output mode has the same form as the input one. As described in Ref.~\cite{molmer_2024}, this assumption leads to the loss of only a few photons from the input mode. In fact, the smaller $\gamma$ is (which means that the coupling is much less than the strength of the pulse) the fewer photons are lost, and the approximation $u(t) = v(t)$ introduces less error.

If the output mode is omitted from consideration (it is allowed to omit the output mode because the dynamics described by Eq.~\eqref{travel-hamiltonian} is unidirectional) and the input mode is in a coherent state, it has been shown in Ref.~\cite{molmer_2020} that the Hamiltonian \eqref{travel-hamiltonian} can be simplified significantly:
\be
\label{travel-hamiltonian-coherent}
\hat{H}(t) = \hat{H}_S(t) + i\sqrt{\gamma}[u(t)\alpha^*(0)\hat{c} - u^*(t)\alpha(0)\hat{c}^\dag],
\ee
where $\alpha(0)$ is the coherent-state amplitude. Comparing with Eq.~\eqref{rabi-hamiltonian}, one can see that these Hamiltonians are equivalent, i.e., the dynamics of the system reduces to simple Rabi oscillations. The Lindblad operator also takes the simple form
\be
\label{travel-lidblad-coherent}
\hat{L}_0(t) = \sqrt{\gamma}\hat{c}.
\ee

Although the scattering problem of a light pulse incident on an atom is significantly simplified using the cavity model with Hamiltonian \eqref{travel-hamiltonian} and Lindblad operator \eqref{travel-lidblad}, there is still room for improvement, because if the system is a two-level atom, almost the whole excitation of the input-cavity goes to the output-cavity with only slight losses due to the skipped output temporal modes. Nevertheless, to solve Eq.~\eqref{travel-hamiltonian} one still needs the full Hilbert space of the input and the output. For example, for an input Fock state with 20 photons, one uses a Hilbert-space dimension of $20*2*20 = 800$, which requires density-matrix operations of size $800*800=640000$. To alleviate this problem, one can use an interaction picture with respect to the input and output cavities, as described in Ref.~\cite{molmer_2023}, by taking the part of the Hamiltonian \eqref{travel-hamiltonian}
\be
\label{travel-hamiltonian-ip_0}
\hat{H}_0(t) = \frac{i}{2}[g^*_u(t)g_v(t)\hat{a}^\dag_u\hat{a}_v - g_u(t)g^*_v(t)\hat{a}_u\hat{a}^\dag_v],
\ee
as the reference Hamiltonian, and the remainder as the interaction. For identical incoming and outgoing modes, the resulting Hamiltonian is
\be
\label{travel-hamiltonian-ip}
\begin{split}
\hat{H}(t) & = \hat{H}_S(t) + i\sqrt{\gamma}[u^*(t)\hat{a}_u^\dag\hat{c} - u(t)\hat{c}^\dag\hat{a}_u] \\
& + \frac{i}{2}\sqrt{\gamma}(\cot{\theta} - \tan{\theta})[u^*(t)\hat{a}_v^\dag\hat{c} - u(t)\hat{c}^\dag\hat{a}_v],
\end{split}
\ee
and the Lindblad operator is
\be
\label{travel-lidblad-ip}
\hat{L}_0(t) = \sqrt{\gamma}\hat{c} - (\cot{\theta} + \tan{\theta})u(t)\hat{a}_v.
\ee
Here the angle $\theta$ is defined by
\be
\label{travel-theta}
\sin^2{\theta(t)} = \int_{0}^{t} dt' |u(t')|^2.
\ee

As described in Ref.~\cite{molmer_2024}, the physical meaning of the "virtual" input ($\hat{a}_u$) and "virtual" output ($\hat{a}_v$) cavity operators changes. The operator $\hat{a}_u$ follows the pulse from the input cavity to the output cavity, while $\hat{a}_v$ acts as an effective buffer whose excitation can contribute to the atomic dynamics at the end of the pulse, reflecting non-Markovian behavior in this representation. We will refer to this buffer further in our presentation as "auxiliary mode".

\section{Analytical Derivation \label{Sec:derivation}}
\subsection{Coherent pulse case}
We start with the simpler case of Eq.~\eqref{travel-hamiltonian-coherent}. For a two-level system in the resonant interaction picture and within the rotating-wave approximation, we take $\alpha(0)$ and $u(t)$ to be real and denote $u(t)$ by $p(t)$ to avoid confusion with the Bloch variable below. The Hamiltonian becomes
\be
\label{2level-coherent-hamiltonian}
\hat{H}(t) = i\sqrt{\gamma}p(t)\alpha(0)[\hat{\sigma}^- - \hat{\sigma}^+],
\ee
while the jump operator has the form
\be
\label{2level-coherent-lidblad}
\hat{L}_0(t) = \sqrt{\gamma}\hat{\sigma}^-.
\ee
Then the equation of motion can be written as
\be
\label{2level-coherent-motion}
\frac{d\rho}{dt} = A + B,
\ee
where
\begin{subequations}
\begin{align}
\label{2level-coherent-motion-A}
A &= -i[\hat{H}, \rho], 
\\
\label{2level-coherent-motion-B}
B &= \hat{L}_0\rho\hat{L}_0^\dag - \frac{1}{2}\{\hat{L}_0^\dag\hat{L}_0, \rho\}.
\end{align}
\end{subequations}

Using the number-state representation in Ref.~\cite{milburn} (p. 99), the matrix elements of $B$ are
\be
\label{B-part}
\begin{split}
B_{mn} & = \bra{m}D[\hat{L}_0]\rho\ket{n} \\
& = \gamma[\sqrt{(m+1)(n+1)}\rho_{m+1,n+1} - \frac{1}{2}(m+n)\rho_{mn}].
\end{split}
\ee

For $A$, we obtain
\be
\label{A-part}
\begin{split}
A_{mn} & = -i\bra{m}[\hat{H}, \rho]\ket{n} \\
& = \sqrt{\gamma}p(t)\alpha(0)\bra{m}[(\hat{\sigma}^- - \hat{\sigma}^+)\rho \\
& - \rho(\hat{\sigma}^- - \hat{\sigma}^+)]\ket{n} \\
& = \sqrt{\gamma}p(t)\alpha(0)(\sqrt{m+1}\rho_{m+1,n} - \sqrt{m}\rho_{m-1,n} \\
& - \sqrt{n}\rho_{m,n-1} + \sqrt{n+1}\rho_{m,n+1}).
\end{split}
\ee
Then the full equation of motion has the form
\be
\label{2level-coherent-motion-full}
\begin{split}
\frac{d\rho_{mn}}{dt} & = \sqrt{\gamma}p(t)\alpha(0)(\sqrt{m+1}\rho_{m+1,n} - \sqrt{m}\rho_{m-1,n} \\
& - \sqrt{n}\rho_{m,n-1} + \sqrt{n+1}\rho_{m,n+1}) \\
& + \gamma[\sqrt{(m+1)(n+1)}\rho_{m+1,n+1} \\
& - \frac{1}{2}(m+n)\rho_{mn}].
\end{split}
\ee
 Written explicitly, these equations are
\be
\label{2level-coherent-motion-full-system}
\begin{split}
\frac{d\rho_{00}}{dt} & = \sqrt{\gamma}p(t)\alpha(0)(\rho_{10} + \rho_{01}) + \gamma\rho_{11}, \\
\frac{d\rho_{11}}{dt} & = -\sqrt{\gamma}p(t)\alpha(0)(\rho_{10} + \rho_{01}) - \gamma\rho_{11}, \\
\frac{d\rho_{01}}{dt} & = \sqrt{\gamma}p(t)\alpha(0)(\rho_{11} - \rho_{00}) - \frac{\gamma}{2}\rho_{01}, \\
\frac{d\rho_{10}}{dt} & = \sqrt{\gamma}p(t)\alpha(0)(\rho_{11} - \rho_{00}) - \frac{\gamma}{2}\rho_{10}.
\end{split}
\ee
For convenience, we introduce the Bloch variables
\be
\label{bloch-basis}
\begin{split}
w & = \rho_{11} - \rho_{00}, \\
u & = \frac{1}{2}(\rho_{10} + \rho_{01}), \\
v & = -\frac{i}{2}(\rho_{10} - \rho_{01}).
\end{split}
\ee

Using $\rho_{11} + \rho_{00} = 1$, we find that $2\rho_{11} = w+1$. The system in Bloch variables becomes
\be
\label{2level-coherent-motion-full-system-bloch}
\begin{split}
\frac{dw}{dt} & = -4\sqrt{\gamma}p(t)\alpha(0)u - \gamma(w+1), \\
\frac{du}{dt} & = \sqrt{\gamma}p(t)\alpha(0)w - \frac{\gamma}{2}u, \\
\frac{dv}{dt} & = - \frac{\gamma}{2}v.
\end{split}
\ee
The equation for $v$ is independent of the rest, while the equations for $w$ and $u$ can be decoupled as follows. From the first equation, we obtain $u$
\be
\label{2level-coherent-motion-u}
u = -\frac{1}{4\sqrt{\gamma}p(t)\alpha(0)}\left[\frac{dw}{dt} + \gamma(w+1)\right].
\ee
 Differentiating the first equation and substituting the second equation and Eq.~\eqref{2level-coherent-motion-u} gives
\be
\label{2level-coherent-motion-w}
\begin{split}
\frac{d^2w}{dt^2} & = \left(\frac{\dot{p}}{p}-\frac{3\gamma}{2}\right)\frac{dw}{dt} \\
& + \left[-4\gamma\alpha(0)^2p^2 + \gamma\left(\frac{\dot{p}}{p}-\frac{\gamma}{2}\right) \right]w  + \gamma\left(\frac{\dot{p}}{p}-\frac{\gamma}{2}\right).
\end{split}
\ee
We are interested in steady-state solutions where $\dot{w} = 0$ and $\ddot{w} = 0$. Such a solution should exist when there are no dissipation channels other than $\hat{L}_0$. We can then derive
\be
\label{2level-coherent-motion-w-eq-solution}
w_{eq} = -\left[\frac{8p^2\alpha(0)^2}{\gamma - 2\dot{p}/p} + 1\right]^{-1}.
\ee
Let us now check the self-consistency by evaluating the time derivative of $w_{eq}$:
\be
\label{2level-coherent-motion-w-eq-solution-consistency}
\dot{w}_{eq} = -8\alpha(0)^2\frac{3p^2(\gamma p -2\dot{p}) - p^3(\gamma\dot{p}-2\ddot{p})}{(8\alpha(0)^2p^3 + \gamma p - 2\dot{p})^2},
\ee
which, under the assumptions that $p(t)$ is finite and $\dot{p},\ddot{p},\gamma \ll \alpha(0)^2$, implies $\dot{w}_{eq} \approx 0$.

Therefore, under such hypotheses, an equilibrium solution $w_{eq}$ exists, as given above. Moreover, under the additional assumption that $\dot{p}/p \ll \gamma$, it simplifies to
\be
\label{2level-coherent-motion-w-eq-solution-simple}
w_{eq} = -\frac{1}{\frac{8}{\gamma}p(t)^2\alpha(0)^2 + 1}.
\ee
If the squared amplitude $\alpha(0)^2$ (i.e., the intensity) is sufficiently large, then at long times $w_{eq} \longrightarrow 0$, i.e., the final state of the system will be a mixed state with equal populations in the ground and excited levels.

\subsection{Fock-state pulse case}
 For a Fock-state pulse, the equation of motion can instead be rearranged to expose the terms governing the dynamics. We again use the resonant interaction picture and the rotating-wave approximation.

As a starting point, we use Eqs.~\eqref{travel-hamiltonian-ip} and \eqref{travel-lidblad-ip} and substitute them into Eq.~\eqref{travel-motion}. First, we calculate
\be
\label{2level-fock-motion-step1}
\begin{split}
L_0^\dag L_0 & = [\sqrt{\gamma}\hat{\sigma}^+ - (\cot{\theta} + \tan{\theta})u(t)\hat{a}_v^\dag] \\
& [\sqrt{\gamma}\hat{\sigma}^- - (\cot{\theta} + \tan{\theta})u(t)\hat{a}_v] \\
& = \gamma\hat{\sigma}^+\hat{\sigma}^- + (\cot{\theta} + \tan{\theta})^2u(t)^2\hat{a}_v^\dag\hat{a}_v \\
& - \sqrt{\gamma}(\cot{\theta} + \tan{\theta})u(t)(\hat{\sigma}^+\hat{a}_v + \hat{\sigma}^-\hat{a}_v^\dag).
\end{split}
\ee
 Next, we calculate
\be
\label{2level-fock-motion-step2}
\begin{split}
L_0 \rho L_0^\dag & = [\sqrt{\gamma}\hat{\sigma}^- - (\cot{\theta} + \tan{\theta})u(t)\hat{a}_v]\rho \\
& [\sqrt{\gamma}\hat{\sigma}^+ - (\cot{\theta} + \tan{\theta})u(t)\hat{a}_v^\dag] \\
& = \gamma\hat{\sigma}^-\rho\hat{\sigma}^+ + (\cot{\theta} + \tan{\theta})^2u(t)^2\hat{a}_v\rho\hat{a}_v^\dag \\
& - \sqrt{\gamma}(\cot{\theta} + \tan{\theta})u(t)(\hat{a}_v\rho\hat{\sigma}^+ + \hat{\sigma}^-\rho\hat{a}_v^\dag).
\end{split}
\ee
 The full dissipative term can then be written as
\be
\label{2level-fock-motion-step3}
\begin{split}
D[L_0]\rho & = D[\sqrt{\gamma}\hat{\sigma}^-]\rho + D[(\cot{\theta} + \tan{\theta})u(t)\hat{a}_v]\rho \\
& + \sqrt{\gamma}(\cot{\theta} + \tan{\theta})u(t)[\frac{1}{2}(\hat{\sigma}^+\hat{a}_v + \hat{a}_v^\dag\hat{\sigma}^-)\rho \\
& + \frac{1}{2}\rho(\hat{\sigma}^+\hat{a}_v + \hat{a}_v^\dag\hat{\sigma}^-) - (\hat{a}_v\rho\hat{\sigma}^+ + \hat{\sigma}^-\rho\hat{a}_v^\dag)].
\end{split}
\ee
For the interaction Hamiltonian, we have
\be
\label{2level-fock-motion-step4}
\begin{split}
-i[\hat{H},\rho] & = \sqrt{\gamma}u(t)[\hat{a}_u^\dag\hat{\sigma}^-\rho - \rho\hat{a}_u^\dag\hat{\sigma}^- - \hat{\sigma}^+\hat{a}_u\rho \\
& + \rho\hat{\sigma}^+\hat{a}_u + \frac{1}{2}(\cot{\theta} - \tan{\theta})(\hat{a}_v^\dag\hat{\sigma}^-\rho \\
& - \rho\hat{a}_v^\dag\hat{\sigma}^- - \hat{\sigma}^+\hat{a}_v\rho + \rho\hat{\sigma}^+\hat{a}_v)].
\end{split}
\ee
Combining the Hamiltonian and dissipative contributions gives
\be
\label{2level-fock-motion-step5}
\begin{split}
\frac{d\rho}{dt} & = \sqrt{\gamma}u(t)(\hat{a}_u^\dag\hat{\sigma}^-\rho - \rho\hat{a}_u^\dag\hat{\sigma}^- - \hat{\sigma}^+\hat{a}_u\rho + \rho\hat{\sigma}^+\hat{a}_u) \\
& + \frac{\sqrt{\gamma}u(t)(\cot{\theta} - \tan{\theta})}{2}(\hat{a}_v^\dag\hat{\sigma}^-\rho - \rho\hat{a}_v^\dag\hat{\sigma}^- \\
& - \hat{\sigma}^+\hat{a}_v\rho + \rho\hat{\sigma}^+\hat{a}_v) \\
& + \frac{\sqrt{\gamma}u(t)(\cot{\theta} + \tan{\theta})}{2}(\hat{\sigma}^+\hat{a}_v\rho + \hat{a}_v^\dag\hat{\sigma}^-\rho \\
& + \rho\hat{\sigma}^+\hat{a}_v + \rho\hat{a}_v^\dag\hat{\sigma}^-) \\
& - \sqrt{\gamma}u(t)(\cot{\theta} + \tan{\theta})(\hat{a}_v\rho\hat{\sigma}^+ + \hat{\sigma}^-\rho\hat{a}_v^\dag) \\
& + D[\sqrt{\gamma}\hat{\sigma}^-]\rho + D[(\cot{\theta} + \tan{\theta})u(t)\hat{a}_v]\rho.
\end{split}
\ee
 Collecting the terms proportional to $\cot{\theta}$ and $\tan{\theta}$, we obtain
\be
\label{2level-fock-motion}
\frac{d\rho}{dt} = A + B + C + D,
\ee
where
\begin{subequations}
\begin{align}
\label{2level-fock-motion-A}
A &= \sqrt{\gamma}u(t)(\hat{a}_u^\dag\hat{\sigma}^-\rho - \rho\hat{a}_u^\dag\hat{\sigma}^- - \hat{\sigma}^+\hat{a}_u\rho + \rho\hat{\sigma}^+\hat{a}_u),
\\
\label{2level-fock-motion-B}
B &= \sqrt{\gamma}u(t)\cot{\theta}(\hat{a}_v^\dag\hat{\sigma}^-\rho + \rho\hat{\sigma}^+\hat{a}_v - \hat{a}_v\rho\hat{\sigma}^+ - \hat{\sigma}^-\rho\hat{a}_v^\dag),
\\
\label{2level-fock-motion-C}
C &= \sqrt{\gamma}u(t)\tan{\theta}(\hat{\sigma}^+\hat{a}_v\rho + \rho\hat{\sigma}^-\hat{a}_v^\dag - \hat{a}_v\rho\hat{\sigma}^+ - \hat{\sigma}^-\rho\hat{a}_v^\dag),
\\
\label{2level-fock-motion-D}
D &= D[\sqrt{\gamma}\hat{\sigma}^-]\rho + D[(\cot{\theta} + \tan{\theta})u(t)\hat{a}_v]\rho.
\end{align}
\end{subequations}
The term $A$ describes Jaynes-Cummings exchange between the virtual input cavity and the two-level system. The term $B$ represents unidirectional population transfer from the two-level system to the virtual output cavity (buffer). The term $C$ represents unidirectional population transfer from the output cavity to the two-level system, and the term $D$ contains the separate dissipation terms for the two-level system and the output cavity.

The important feature is the interplay between $B$ and $C$: $B$ is weighted by $\cot{\theta}$ and $C$ by $\tan{\theta}$. As will be shown in the simulation results in the next section, $\cot{\theta}$ is large at the beginning of the pulse, while $\tan{\theta}$ is large at the end of the pulse. This suggests why population growth of the two-level system is observed at the end of the pulse. In fact, the output cavity in the interaction picture plays the role of a buffer in which excitation is temporarily stored and can contribute to the later recovery of the atomic oscillations.

\section{Numerical results \label{Sec:results}}
\begin{figure}[tbph]
\begin{tabular}{c}
\includegraphics[width=1\columnwidth]{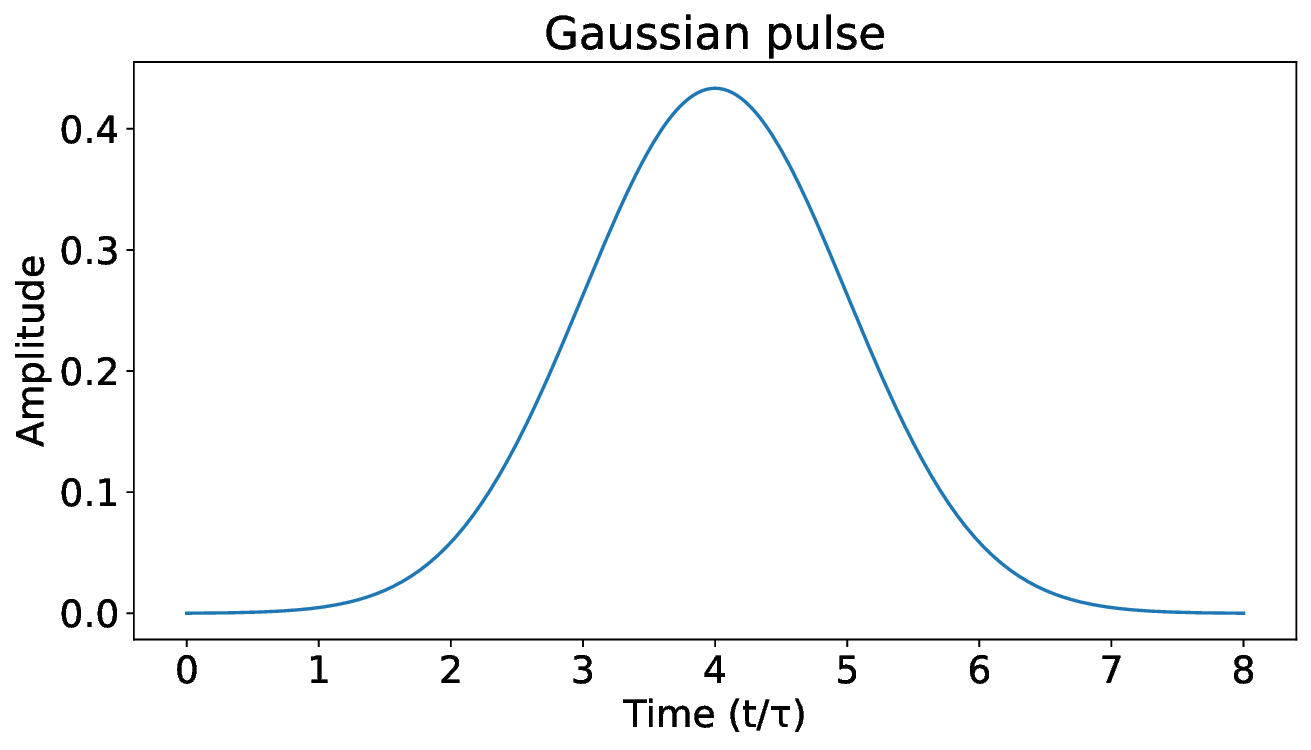}
 \\
    \includegraphics[width=1\columnwidth]{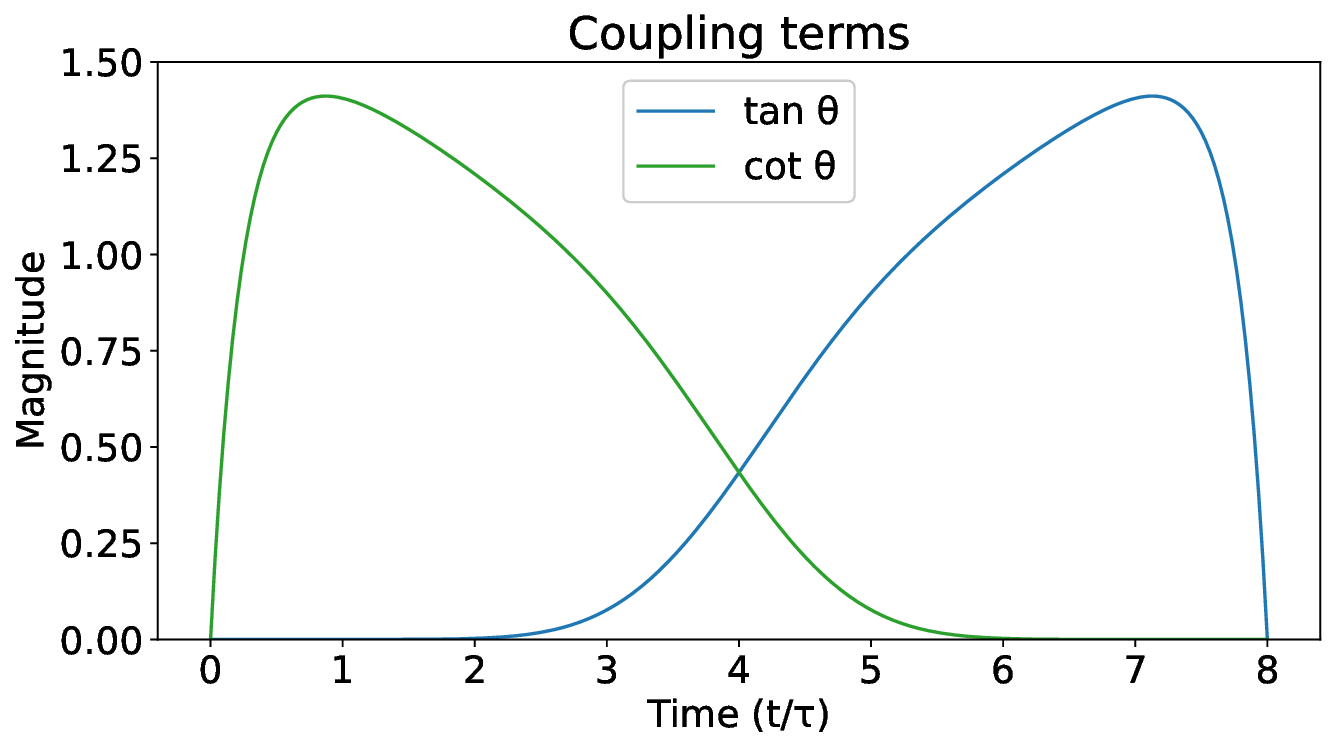}
\end{tabular}
    \caption{Top: Form of the input Gaussian pulse. 
    Bottom: $\cot{\theta}$ and $\tan{\theta}$ as functions of time.
    }
    \label{fig-gaussian-pulse}
\end{figure}

The QuTiP simulations for coherent and Fock pulses use a Gaussian pulse of the form given in Ref.~\cite{molmer_2020}
\be
\label{gaussian-pulse}
u(t) = \frac{1}{\sqrt{\tau}\pi^{1/4}}e^{-\frac{(t-t_p)^2}{2\tau^2}}.
\ee
It is shown in Fig. \ref{fig-gaussian-pulse}(top).
The next plot in Fig. \ref{fig-gaussian-pulse}(bottom) displays how $\cot{\theta}$ and $\tan{\theta}$ evolve with time.
 It shows that term $B$ in Eq.~\eqref{2level-fock-motion-B} contributes at the beginning of the evolution, while term $C$ in Eq.~\eqref{2level-fock-motion-C} contributes at its end.

\begin{figure}[tbph]
    \includegraphics[width=1\columnwidth]
    {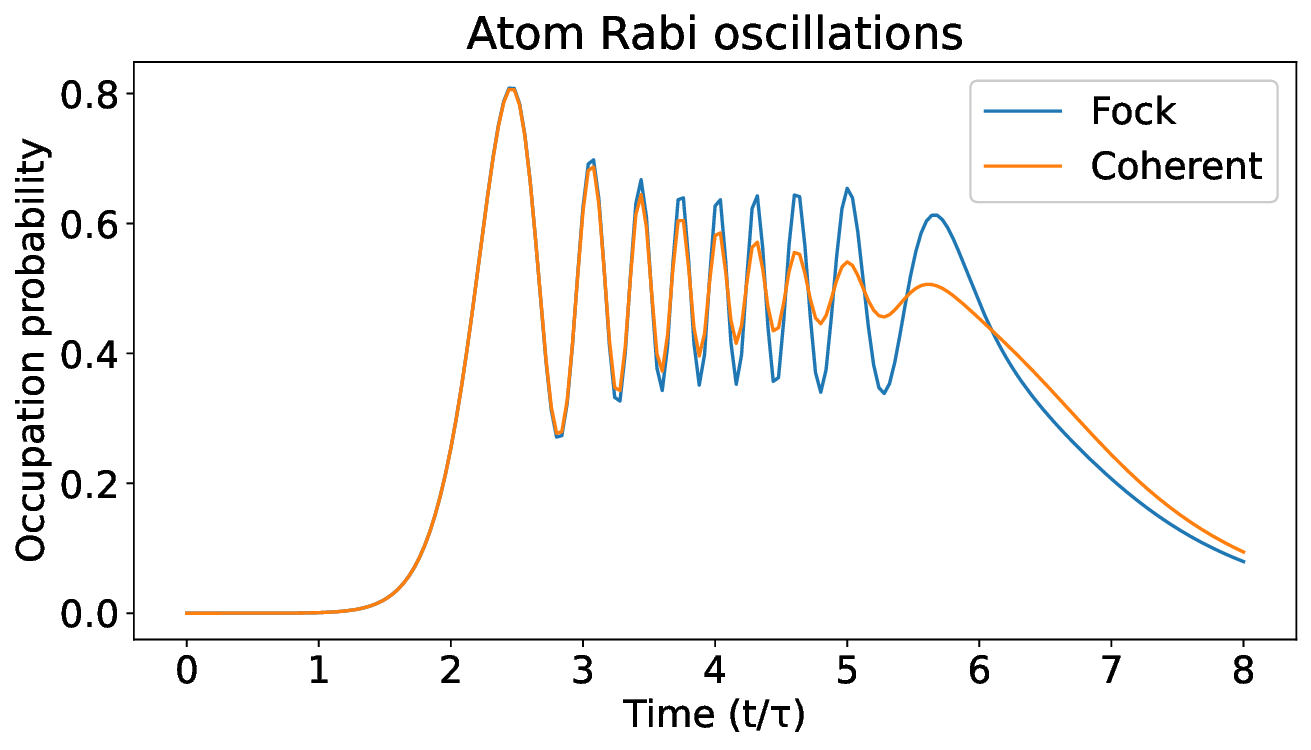}
    \caption{Comparison of driven Fock-state and coherent-state oscillations.}
    \label{fig-fock-and-coherent-oscillations}
\end{figure}

 Simulations of the Hamiltonian \eqref{travel-hamiltonian-ip} with $\gamma = 1/\tau$ for Fock and coherent states illustrate the behavior discussed in Section \ref{Sec:derivation}. The coherent-state pulse experiences exponential decay to a mixed state with equal populations, while the Fock-driven oscillations recover near the end of the pulse, reminiscent of oscillations revival, as suggested by the preceding analysis (see Fig. \ref{fig-fock-and-coherent-oscillations}).

It is important to note that if the output-cavity terms are removed from Eq.~\eqref{travel-hamiltonian-ip}, then the Fock-state behavior closely resembles the coherent-state behavior, which supports the role of the population stored in the output cavity.

 A further simulation shows that if, for coherent states, we use the Hamiltonian \eqref{travel-hamiltonian-coherent} instead of Eq.~\eqref{travel-hamiltonian-ip}, the result is unchanged, as expected.

\begin{figure}[tbph]
\begin{tabular}{c}
\includegraphics[width=1\columnwidth]    {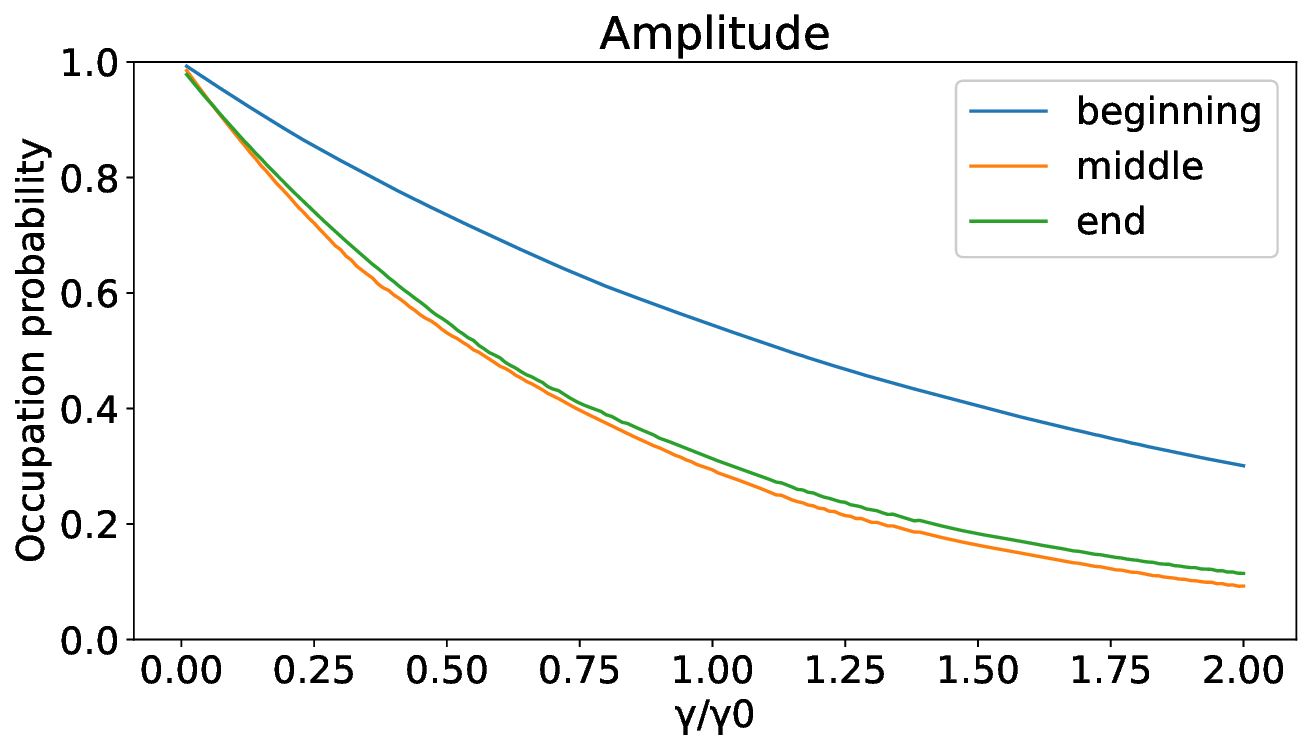} \\
    \includegraphics[width=1\columnwidth]    {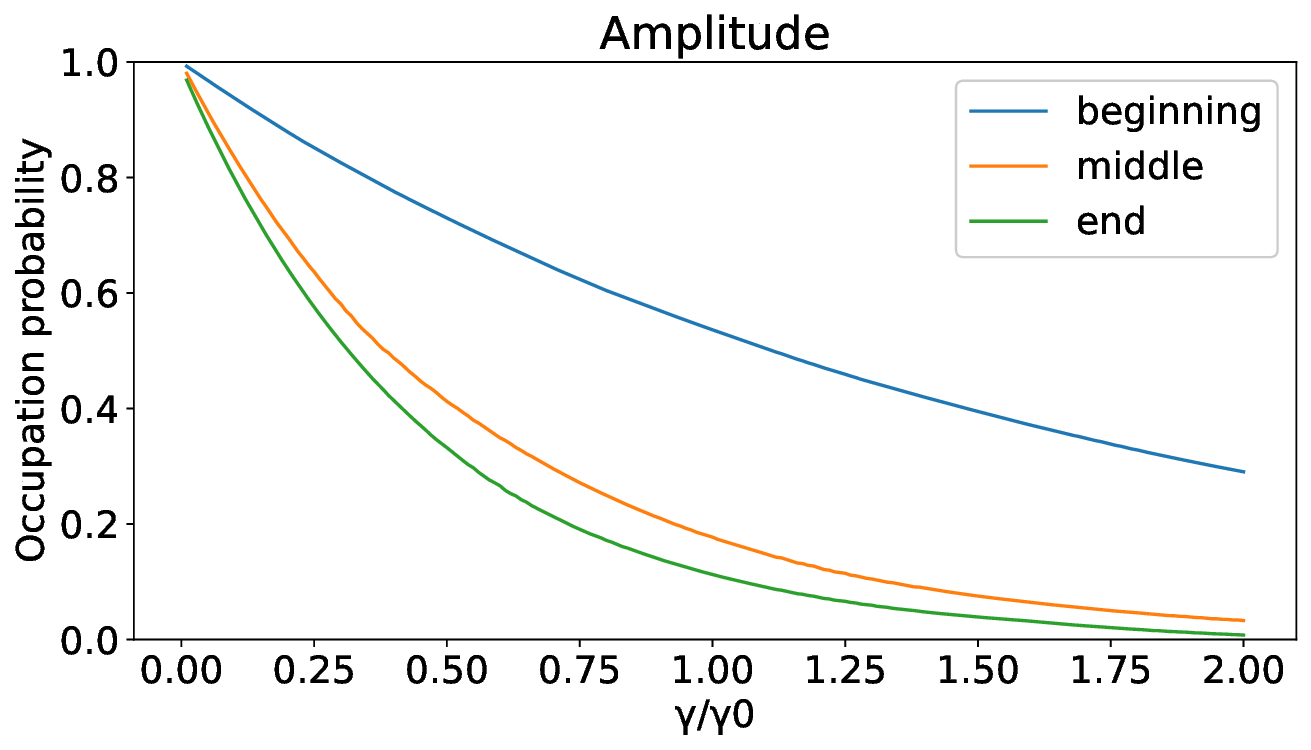}
    \end{tabular}
    \caption{Top: Dependence on $\gamma$ for three parts of a Fock pulse.
    Bottom: Dependence on $\gamma$ for three parts of a coherent pulse.}
    \label{fig-fock-of-gamma}
\end{figure}

 In Fig.~\ref{fig-fock-of-gamma}, the occupation probability as a function of $\gamma$ is shown for Fock and coherent states. 
 Here $\gamma_0=1/\tau$, and three curves refer to the beginning, middle, and end of the pulse. The beginning amplitude is defined as the maximum population difference in the interval $2 < t/\tau < 3.5$, the middle amplitude is defined as the maximum population difference in the interval $3.5 < t/\tau < 4.5$, and the end amplitude is defined as the maximum population difference in the interval $4.5 < t/\tau < 5.5$ (see Fig. \ref{fig-fock-and-coherent-oscillations}).
 
The figures clearly show that for a Fock state the lowest amplitude in the examples shown is in the middle of the pulse, while for a coherent state the lowest amplitude in the examples shown is at the end of the pulse.

\section{Analogy with STIRAP \label{Sec:stirap}}
\subsection{Three-level system description}
\begin{figure}[tbph]
    \begin{tabular}{rrr}
        \includegraphics[width=0.985\columnwidth]{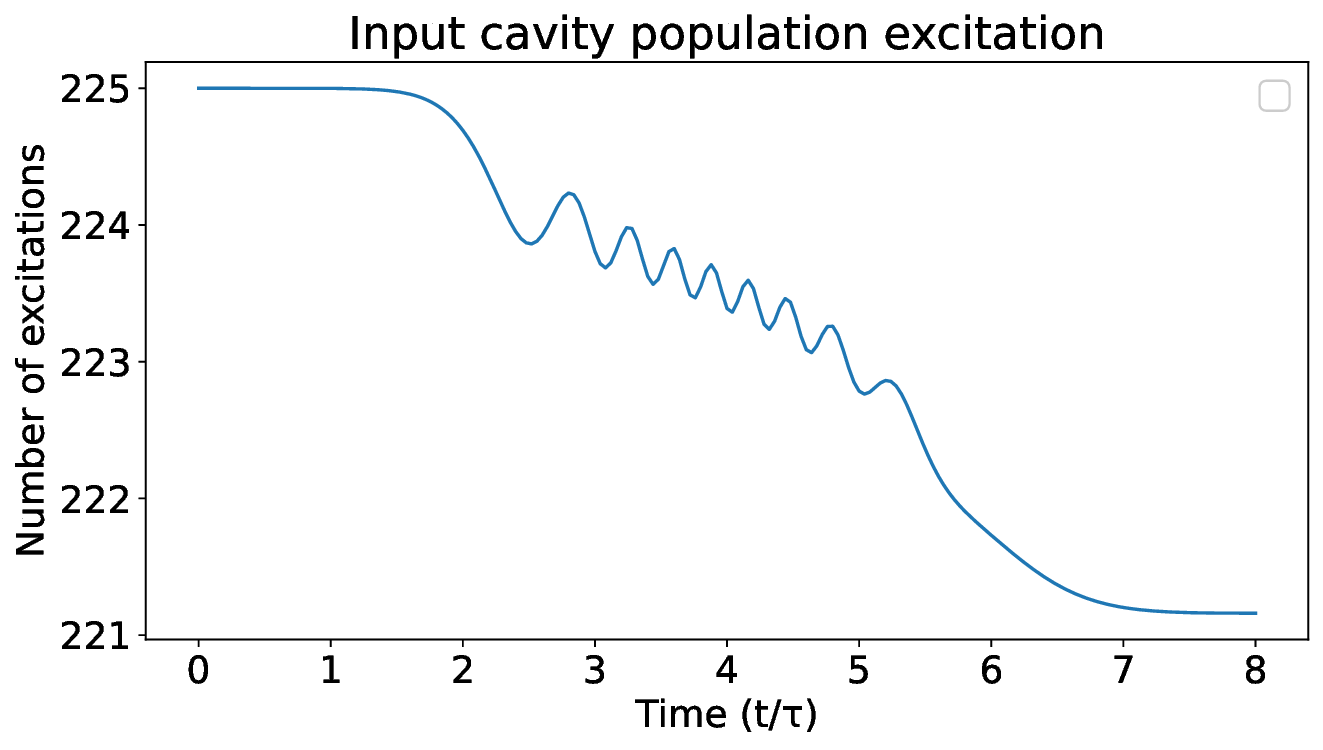} \\
        \includegraphics[width=1\columnwidth]{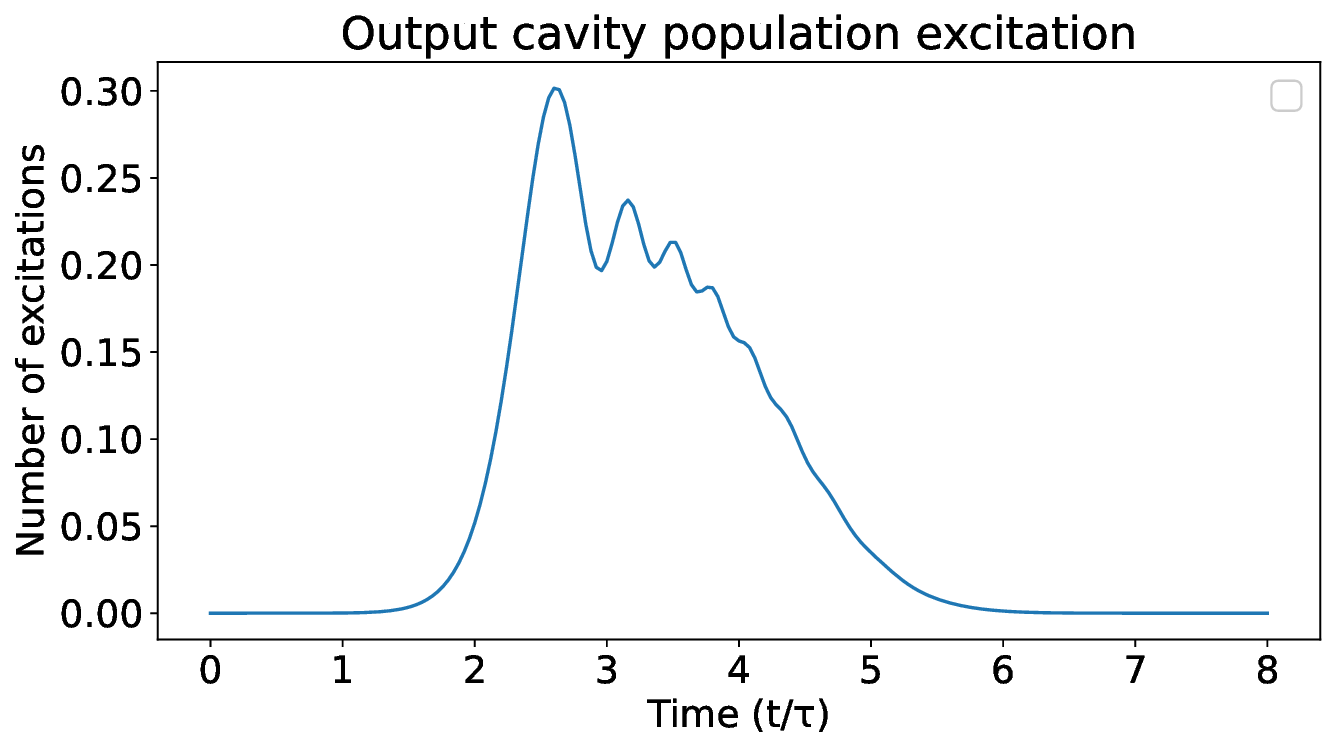} \\
        \includegraphics[width=0.985\columnwidth]{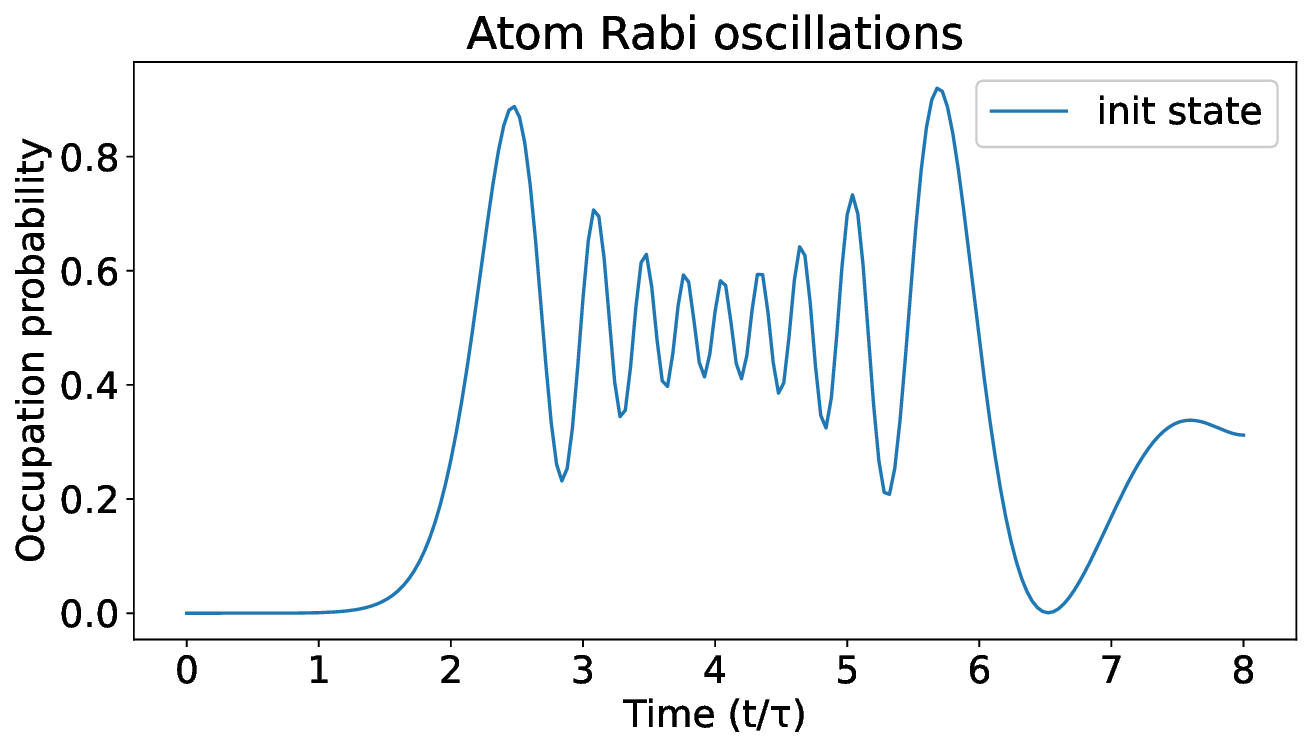}
    \end{tabular}
    \caption{Fock-state pulse dynamics. Top: mean input-cavity occupation; middle: mean output-cavity (auxiliary-mode) occupation, including dissipation, for an initial input state with 225 photons; bottom: atomic population calculated without the Lindblad term.}
    \label{fig-fock-pulse-dynamics}
\end{figure}

As discussed in the introduction, the interaction picture defined by Eq.~\eqref{travel-hamiltonian-ip_0} can concentrate the dynamics near a small set of number states. Top and middle panels of Fig.~\ref{fig-fock-pulse-dynamics} show the mean occupations of the two virtual modes for the pulse in Fig.~\ref{fig-gaussian-pulse}, initially in a Fock state with 225 photons. This photon number was chosen to make the recovery more visible, without assuming that such a state can presently be prepared.

 The mean input-mode occupation changes by only a few photons, while the auxiliary-mode occupation remains below one. These means suggest a small numerical basis, but do not by themselves establish that only five and two number states are populated, respectively. However in the limit of large photon number and small $\gamma$, the relevant input cavity modes can be approximated just 2. To examine the coherent coupling structure, we also omit the first dissipator associated with Eq.~\eqref{travel-lidblad-ip}. The second term of the jump operator describes loss from the auxiliary mode, whose transient occupation is shown in Fig.~\ref{fig-fock-pulse-dynamics}(middle). Omitting the full dissipator makes the recovery more pronounced, as the comparison of Figs.~\ref{fig-fock-pulse-dynamics}(bottom) and \ref{fig-fock-and-coherent-oscillations} shows. We use this simpler model for a qualitative comparison, without identifying its recovery amplitude with that of the dissipative system.

 A three-state description is obtained by projecting Eq.~\eqref{travel-hamiltonian-ip} onto $\ket{0}=\ket{n,g,0}$, $\ket{1}=\ket{n-1,e,0}$, and $\ket{2}=\ket{n-1,g,1}$, where the entries denote the input-mode occupation, the atomic state, and the auxiliary-mode occupation. This gives
\be
\label{travel-hamiltonian-ip-js}
\begin{split}
\hat{H}(t) & = i\sqrt{n\gamma}u(t)[\ket{0}\bra{1} - \ket{1}\bra{0}] \\
& + \frac{i}{2}\sqrt{\gamma}p(t)u(t)[\ket{2}\bra{1} - \ket{1}\bra{2}].
\end{split}
\ee
 The factor $\sqrt n$ comes from the input-mode absorption matrix element. We take $u(t)$ real and define $p(t)=\cot\theta-\tan\theta$. For $n=1$, the three-state subspace is invariant under the Hamiltonian. For $n>1$, however, $\ket{2}$ also couples to $\ket{n-2,e,1}$, and the full fixed-excitation sector has dimension $2n+1$. A mode-state correspondence or Jordan-Schwinger mapping \cite{jordan_1935,schwinger_1952,huang_2024} does not remove these additional states. Equation~\eqref{travel-hamiltonian-ip-js} is therefore a three-state approximation for $n>1$.

\begin{figure}[tbph]
    \includegraphics[width=1\columnwidth]
    {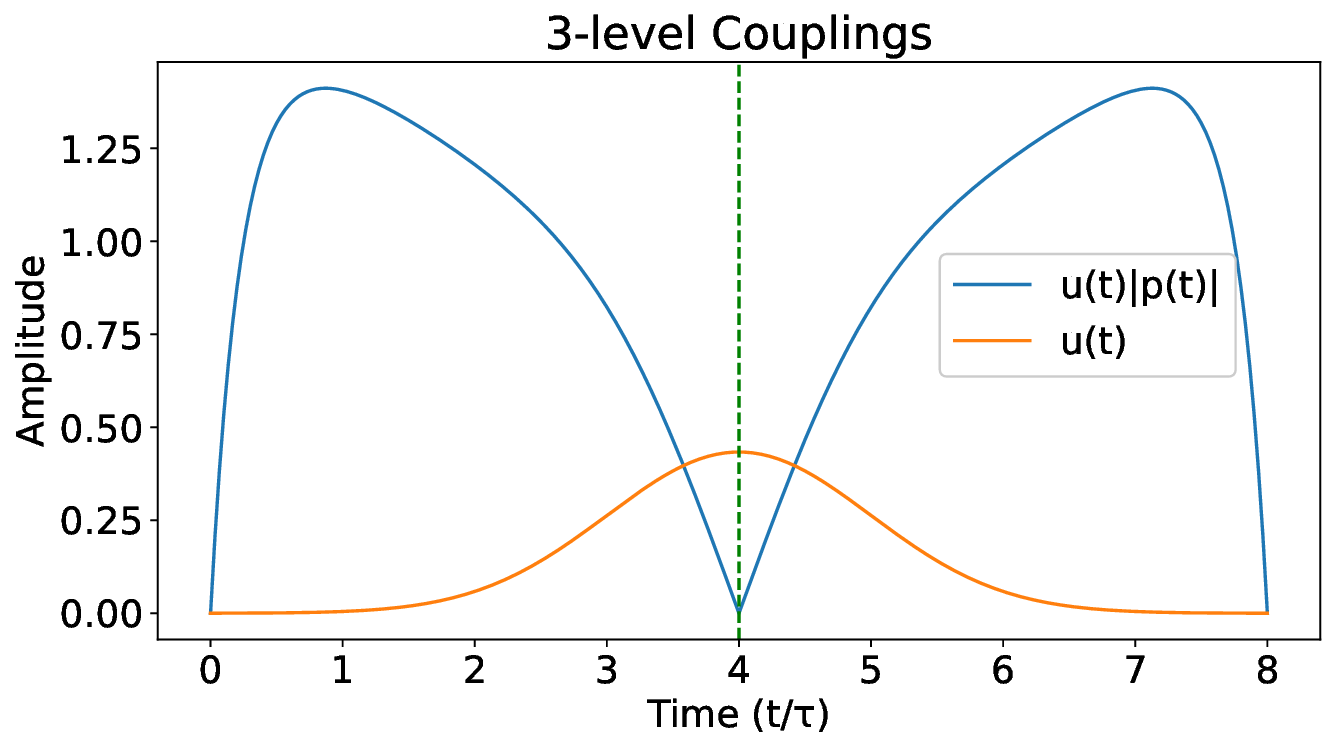}
    \caption{Two mirrored STIRAP-like sequences for couplings in Eq.~\eqref{travel-hamiltonian-ip-js}.}
    \label{fig-stirap_couplings}
\end{figure}

 Figure~\ref{fig-stirap_couplings} shows the temporal factors $u(t)$ and $u(t)|p(t)|$ entering these couplings. The constant factors in Eq.~\eqref{travel-hamiltonian-ip-js} are not included, and the absolute value hides the sign change of $p(t)$.

In the first half of the pulse, the coupling between $\ket{1}$ and $\ket{2}$ dominates first; the coupling between $\ket{0}$ and $\ket{1}$ then becomes relatively stronger. This is the counterintuitive ordering used in STIRAP to transfer population from $\ket{0}$ to $\ket{2}$, as reviewed in Ref.~\cite{Vitanov2017STIRAP}. The second half reverses the ordering. The resemblance suggests a connection with the initial reduction and later recovery of the atomic oscillations. Establishing adiabatic transfer, however, requires a dark state and an adiabaticity check, rather than pulse ordering alone.

Adiabatic transfer between two quantized modes coupled through a two-level atom was introduced in Ref.~\cite{Mattinson2001transfer}. The many-photon dark-state construction used below follows the approach of Ref.~\cite{Larson2005cavity}. The full coherent Hamiltonian also has an exact many-photon dark state. Set $G_u=\sqrt\gamma\,u(t)$, $G_v=\sqrt\gamma\,u(t)p(t)/2$, and $G=(G_u^2+G_v^2)^{1/2}$. For real couplings and $G\ne0$, define
\be
\label{stirap-bright-dark}
\begin{split}
\hat b&=(G_u\hat a_u+G_v\hat a_v)/G,\\
\hat d^\dag&=(G_v\hat a_u^\dag-G_u\hat a_v^\dag)/G.
\end{split}
\ee
Then $\hat H=iG(\hat b^\dag\hat\sigma^-\hat\sigma^+\hat b)$ and $[\hat b,\hat d^\dag]=0$. Hence
\be
\label{stirap-many-photon-dark}
\ket{D_n(t)}=\frac{[\hat d^\dag(t)]^n}{\sqrt{n!}}\ket{0,g,0},
\quad\hat H\ket{D_n}=0.
\ee
For $n=1$, this is the usual three-state dark state; for $n>1$, it contains $n+1$ field-number configurations. A continuous angle $\beta$ defined by $\cos\beta=G_v/G$ and $\sin\beta=G_u/G$ gives $\tan\beta=2/p$. As $p$ passes from large positive to large negative values, the dark state connects the input mode to the auxiliary mode and back, up to a phase. This establishes the double-STIRAP connection at the Hamiltonian level. Along this ideal path,
\be
\label{stirap-dark-population}
P_e^{(D)}=0,\qquad
\langle\hat N_v\rangle_D=\frac{4n}{p^2+4}.
\ee
At $p=0$, the auxiliary mode would contain all $n$ photons. Its maximum mean occupation of about $0.3$ in Fig.~\ref{fig-fock-pulse-dynamics}(middle), for $n=225$, shows that the displayed dynamics does not follow this ideal path. The analogy therefore supplies a reference for the oscillatory dynamics, rather than proof of adiabatic transfer.

\subsection{Adiabaticity considerations}
 The STIRAP analogy provides an adiabaticity diagnostic for the three-state model \cite{vitanov,shore}. For $n>1$, this diagnostic must be distinguished from adiabatic following in the full excitation sector.

 After constant rephasings of the states, Eq.~\eqref{travel-hamiltonian-ip-js} takes the matrix form
\be
\label{travel-hamiltonian-ip-js-matrix}
\hat{H}(t) = 
\begin{bmatrix}
0 & \sqrt{n\gamma}u(t) & 0\\
\sqrt{n\gamma}u(t) & 0 & \sqrt{\gamma}p(t)u(t)/2\\
0 & \sqrt{\gamma}p(t)u(t)/2 & 0
\end{bmatrix}.
\ee
 The Hamiltonian of a resonant three-level system is
\be
\label{three-level-system}
\hat{H}(t) =
\begin{bmatrix}
0 & \Omega_p(t)/2 & 0\\
\Omega_p(t)/2 & 0 & \Omega_s(t)/2\\
0 & \Omega_s(t)/2 & 0
\end{bmatrix},
\ee
which gives the associations
\be
\label{opmega-p}
\Omega_p(t) = 2\sqrt{n\gamma}u(t),\quad
\Omega_s(t) = \sqrt{\gamma}u(t)p(t).
\ee

 The corresponding mixing angle and generalized Rabi frequency are
\begin{subequations}
\begin{align}
\label{tg-chi}
\tan{\chi}(t) &= \frac{\Omega_p(t)}{\Omega_s(t)} = \frac{2\sqrt{n}}{p(t)},
\\
\label{omega-total}
\Omega(t) &= \sqrt{\Omega_p(t)^2 + \Omega_s(t)^2} = \sqrt{\gamma}u(t)\sqrt{4n + p(t)^2},
\\
\label{chi-angle-prim}
\chi'(t) &= 
-\frac{2\sqrt{n}p'(t)}{p(t)^2+4n}.
\end{align}
\end{subequations}

 Up to a numerical factor of order unity, the local adiabaticity condition is
\be
\label{adiabat-condition-general}
|\chi'(t)| \ll \Omega(t),
\ee
which gives
\be
\label{adiabat-condition}
\frac{2\sqrt{n}|p'(t)|}{p(t)^2+4n} \ll \sqrt{\gamma}u(t)\sqrt{4n + p(t)^2}.
\ee
 Simplification gives
\be
\label{adiabat-condition-simple}
|p'(t)| \ll \frac{\sqrt\gamma\,u(t)}{2\sqrt{n}}[4n + p(t)^2]^{3/2}.
\ee

\begin{figure}[tbph]
\begin{tabular}{c}
    \includegraphics[width=1\columnwidth]{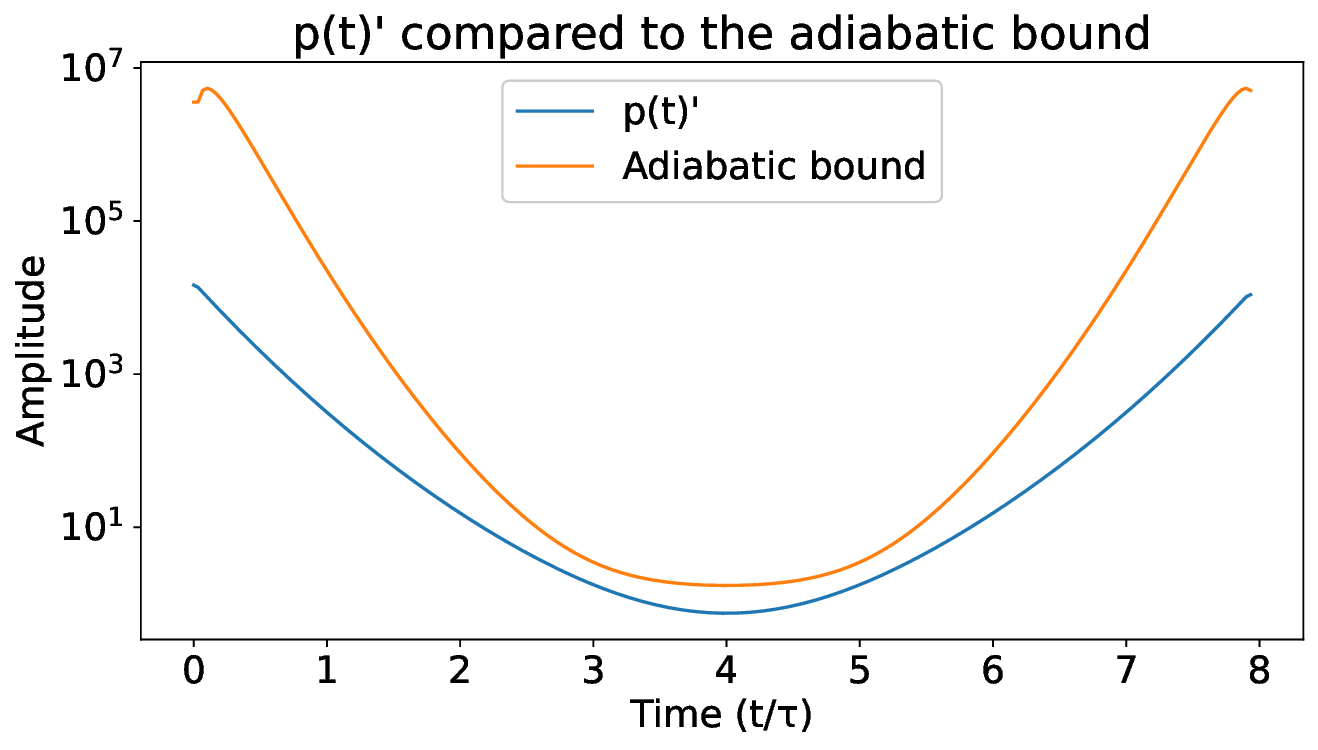} \\
    \includegraphics[width=1\columnwidth]{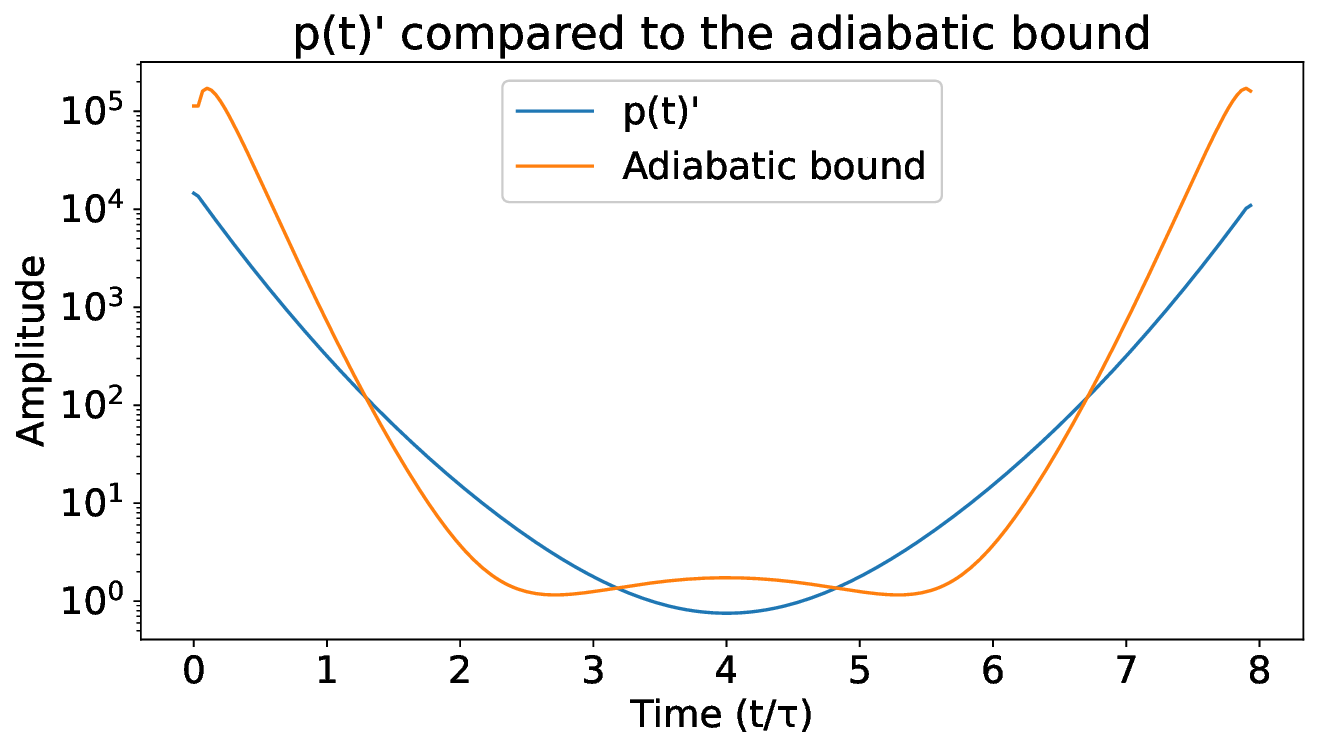}
\end{tabular}
    \caption{LHS vs. RHS of Eq.~\eqref{adiabat-condition-simple}. Top: $n=2$ and $\gamma=1.5/\tau$. Bottom: $n=20$ and $\gamma=0.15/\tau$.}
    \label{fig-adiabatic-bounds}
\end{figure}

Figure~\ref{fig-adiabatic-bounds} shows the original three-state adiabaticity diagnostic for two values of $n$ and $\gamma$. 
 The plots suggest that increasing $n$ and decreasing $\gamma$ at fixed $n\gamma$ broadens the interval in which the three-state diagnostic fails. Perfect following of a dark state would suppress atomic excitation; appreciable Rabi oscillations therefore indicate departure from that path.

For the exact state $\ket{D_n}$, the nearest bright-state gaps have magnitude $G$, while the corresponding derivative couplings have magnitude $\sqrt{n/2}\,|\dot\beta|$. A local adiabaticity diagnostic, apart from a numerical factor of order unity, is therefore
\be
\label{stirap-exact-adiabaticity}
\frac{\sqrt n\,|\dot\beta|}{G}
=\frac{4\sqrt n\,|\dot p|}{\sqrt\gamma\,u(t)(p^2+4)^{3/2}}\ll1.
\ee
The angle must be continued smoothly through $p=0$. This condition differs from the three-state diagnostic for $n>1$ and becomes more demanding as $n$ increases at fixed $n\gamma$. Moreover, $\ket{D_n}$ is not generally dark to the jump operator \eqref{travel-lidblad-ip}: its auxiliary-mode population can decay even though the atom remains unexcited. Adiabaticity of the Hamiltonian alone is therefore insufficient for low-loss transfer.

\subsection{Full Hamiltonian description \label{full-hamiltonian-description}}
 We now return to the original Hamiltonian \eqref{travel-hamiltonian} and compare its evolution with a normal-mode reference that matches the initial state. The construction below diagonalizes a linear three-oscillator Hamiltonian and also applies to the single-excitation matrix of the atom-field system. For a two-level atom with more than one excitation, it does not generally yield an exact instantaneous eigenstate. We omit the dissipative term associated with Eq.~\eqref{travel-lidblad} in this comparison.

 For three bosonic modes with real, possibly time-dependent couplings, the Hamiltonian,
\be
\label{3-mode-hamiltonian}
\begin{split}
\hat{H}(t) & = \sum_k{\omega\hat{a}^\dag_k\hat{a}_k} + i[\epsilon_{12}(t) \hat{a}^\dag_1 \hat{a}_2 + \epsilon_{13}(t) \hat{a}^\dag_1 \hat{a}_3 \\
& + \epsilon_{23}(t) \hat{a}^\dag_2 \hat{a}_3 - h.c.],
\end{split}
\ee
can be diagonalized at each instant as follows (assuming equal mode frequencies):
\be
\label{3-mode-hamiltonian-diagonal}
\hat{H}(t) = \sum_k{\omega_k\hat{A}^\dag_k\hat{A}_k},
\ee
with
\be
\label{3-mode-hamiltonian-diagonal-elements}
\begin{split}
& \omega_1 = \omega, \quad \omega_2 = \omega - \epsilon, \quad \omega_3 = \omega + \epsilon, \\
& \epsilon = \sqrt{\epsilon_{12}^2 + \epsilon_{13}^2 + \epsilon_{23}^2},
\end{split}
\ee
 where the operators $\hat A_k$ are obtained from the eigenvectors of the coefficient matrix
\be\label{3-mode-hamiltonian-matrix}
\hat{H}(t) = \begin{bmatrix}
\hat{a}^\dag_1 & \hat{a}^\dag_2 & \hat{a}^\dag_3
\end{bmatrix}
 \begin{bmatrix}
\omega & i\epsilon_{12} & i\epsilon_{13}\\
-i\epsilon_{12} & \omega & i\epsilon_{23}\\
-i\epsilon_{13} & -i\epsilon_{23} & \omega
\end{bmatrix}\begin{bmatrix}
\hat{a}_1\\
\hat{a}_2\\
\hat{a}_3
\end{bmatrix}.
\ee
 With the associations
\be
\label{3-mode-hamiltonian-associations}
\begin{split}
& \hat{a}_1 = \hat{a}_u, \quad \hat{a}_2 = \hat{a}_v, \quad \hat{a}_3 = \hat{c}, \\
& \epsilon_{13} = \frac{\sqrt{\gamma}g_u}{2}, \quad \epsilon_{23} = -\frac{\sqrt{\gamma}g_v}{2}, \quad \epsilon_{12} = \frac{g_u g_v}{2},
\end{split}
\ee
 one normal-mode combination has the form
\be
\label{3-mode-hamiltonian-eigenstate}
\begin{split}
\hat{A}_1(t) & = -\frac{\sqrt{\gamma}g_v}{\sqrt{\gamma(g_u^2 + g_v^2) + g_u^2 g_v^2}}\hat{a}_u \\
& -\frac{\sqrt{\gamma}g_u}{\sqrt{\gamma(g_u^2 + g_v^2) + g_u^2 g_v^2}}\hat{a}_v \\
& +\frac{g_u g_v}{\sqrt{\gamma(g_u^2 + g_v^2) + g_u^2 g_v^2}}\hat{c}.
\end{split}
\ee

 In the early-time asymptotic limit, $g_u/g_v\to0$ and $g_u/\sqrt\gamma\to0$, so that, up to a phase,
\be
\label{3-mode-hamiltonian-eigenstate-limit}
\hat{A}_1(0) = \hat{a}_u.
\ee

\begin{figure}[tbph]
\begin{tabular}{c}
    \includegraphics[width=1\columnwidth]{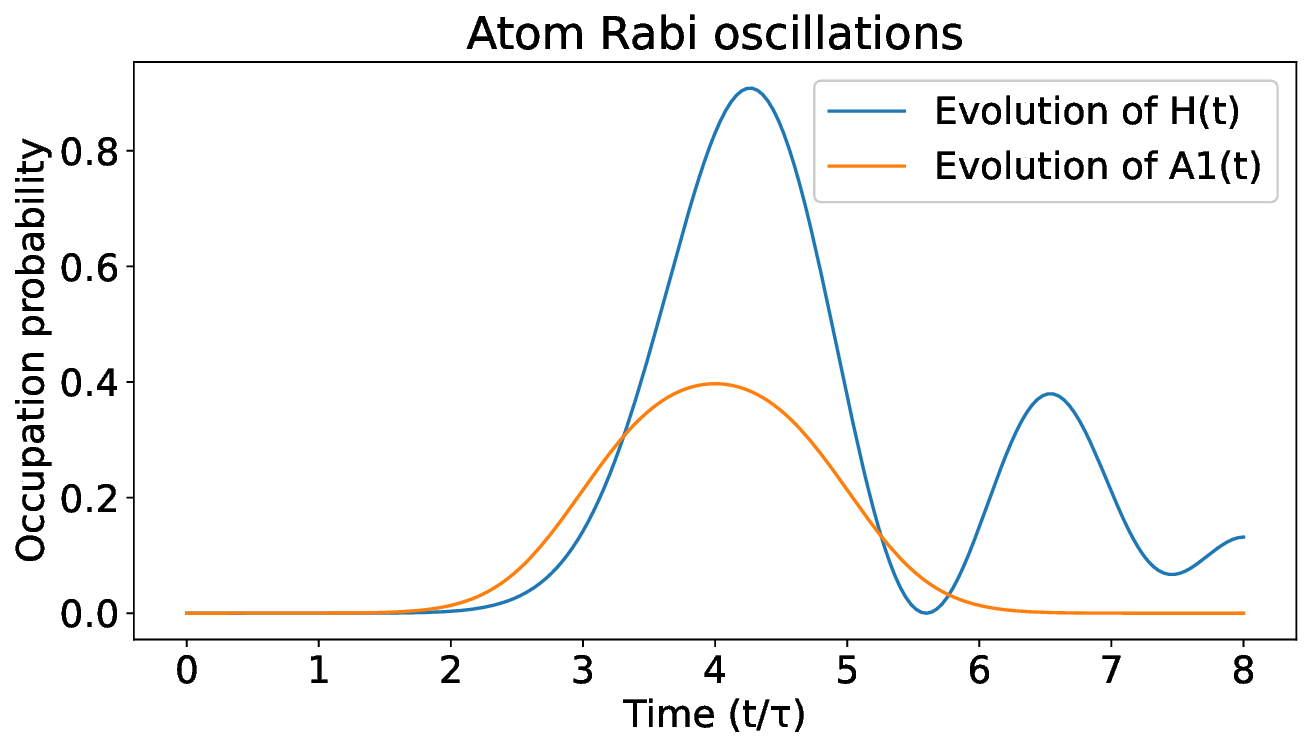} \\
    \includegraphics[width=1\columnwidth]{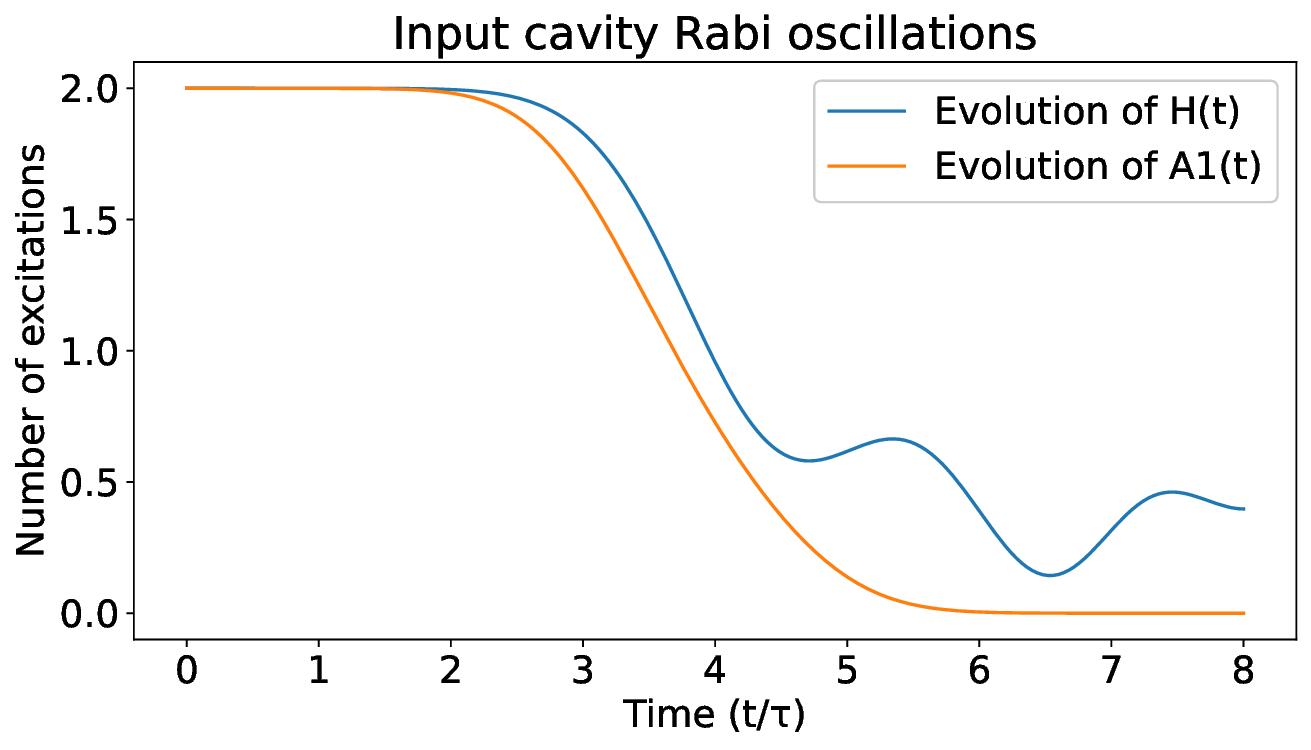} \\
    \includegraphics[width=1\columnwidth]{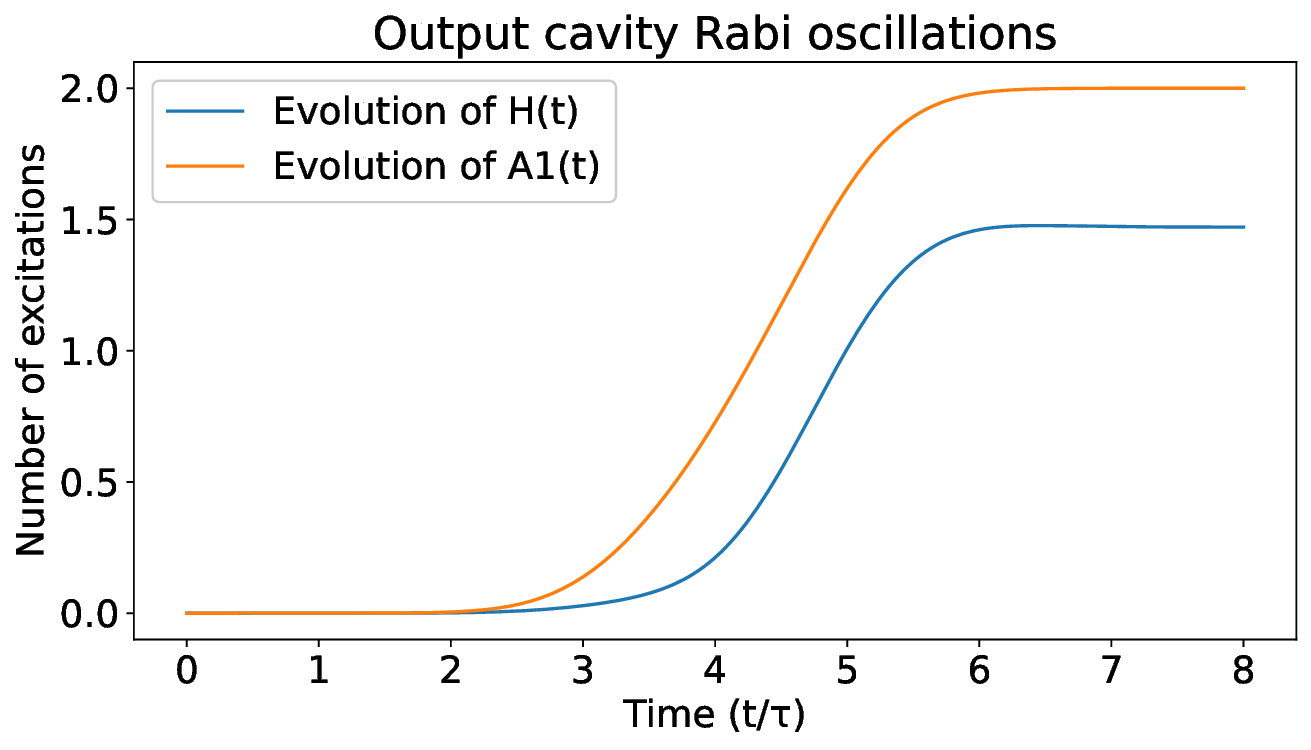}
\end{tabular}
    \caption{Population dynamics for $n=2$ and $\gamma=1.5/\tau$. Top: atomic population. Middle: input-cavity population. Bottom: output-cavity population.}
    \label{fig-adiabat_non_adiabat_populations_2}
\end{figure}

 The initial Fock state can therefore be represented by applying either $\hat a_u^\dag$ or $\hat A_1^\dag$ $n$ times, with appropriate normalization. At later times, the normalized state proportional to $[\hat A_1^\dag(t)]^n\ket{0,g,0}$ provides a reference for comparison with the evolution under Eq.~\eqref{travel-hamiltonian}. For the atom, $\hat c=\hat\sigma^-$ is a spin lowering operator, with $[\hat c,\hat c^\dag]=-\hat\sigma_z$ and $(\hat c^\dag)^2=0$. Restricting the atomic Hilbert space to two dimensions enforces saturation but does not make the bosonic normal-mode construction an exact multiphoton eigenstate. The following figures compare this reference state with the Hamiltonian dynamics. Figure~\ref{fig-adiabat_non_adiabat_populations_2} compares the predicted evolution under the original Hamiltonian $H(t)$ with that of the adiabatic state $A_1(t)$ for the atom (top), input cavity (middle), and output cavity (bottom).

\begin{figure}[tbph]
    \includegraphics[width=1\columnwidth]
    {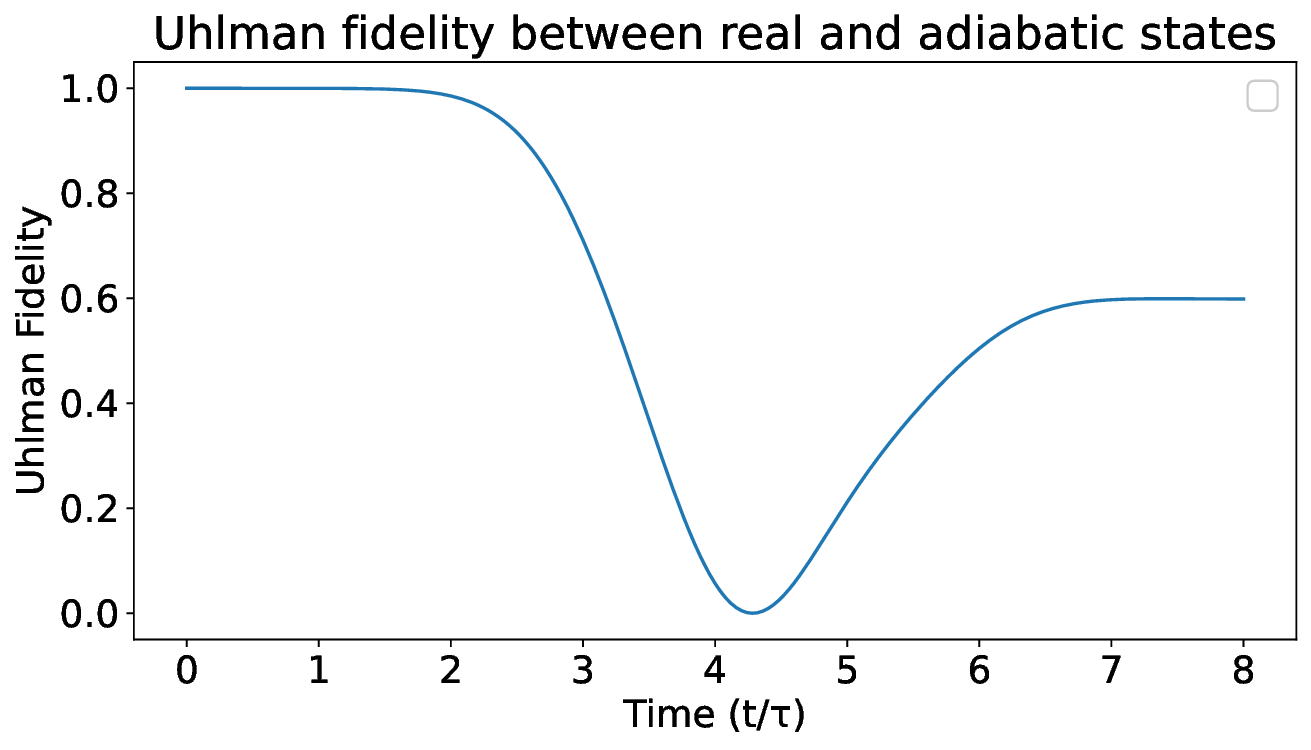}
    \caption{Fidelity of states for $n=2$ and $\gamma=1.5/\tau$.}
    \label{fig-adiabat_non_adiabat_fidelity_2}
\end{figure}
 
 Finally, Fig. \ref{fig-adiabat_non_adiabat_fidelity_2} gives the Uhlmann fidelity \cite{hashim_2025} between the states produced with the two methods. The reference agrees closely with the true state initially, differs during approximately $2<t/\tau<6$, and approaches it again for $6<t/\tau<8$. Because the reference is not generally an eigenstate for $n>1$, this agreement alone does not establish adiabatic following. The final fidelity remains below unity. For the displayed examples, the final fidelity is largest near values of $\gamma n$ corresponding to an effective $2\pi$ atomic rotation. Away from these values, the fidelity can be considerably smaller. This dependence is consistent with phase accumulation during the intermediate atomic oscillations. The time interval of disagreement is also comparable to that indicated by Fig.~\ref{fig-adiabatic-bounds}(top).

\begin{figure}[tbph]
\begin{tabular}{c}
    \includegraphics[width=1\columnwidth]{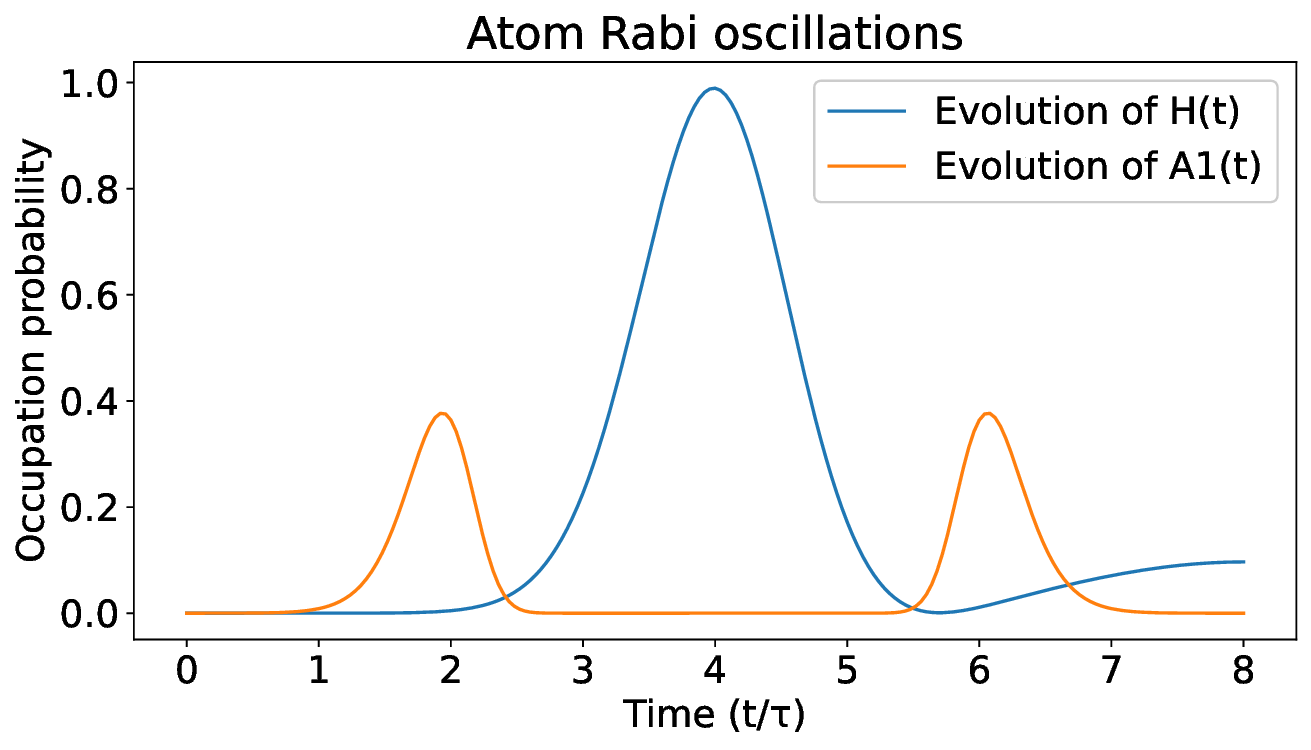} \\
    \includegraphics[width=1\columnwidth]{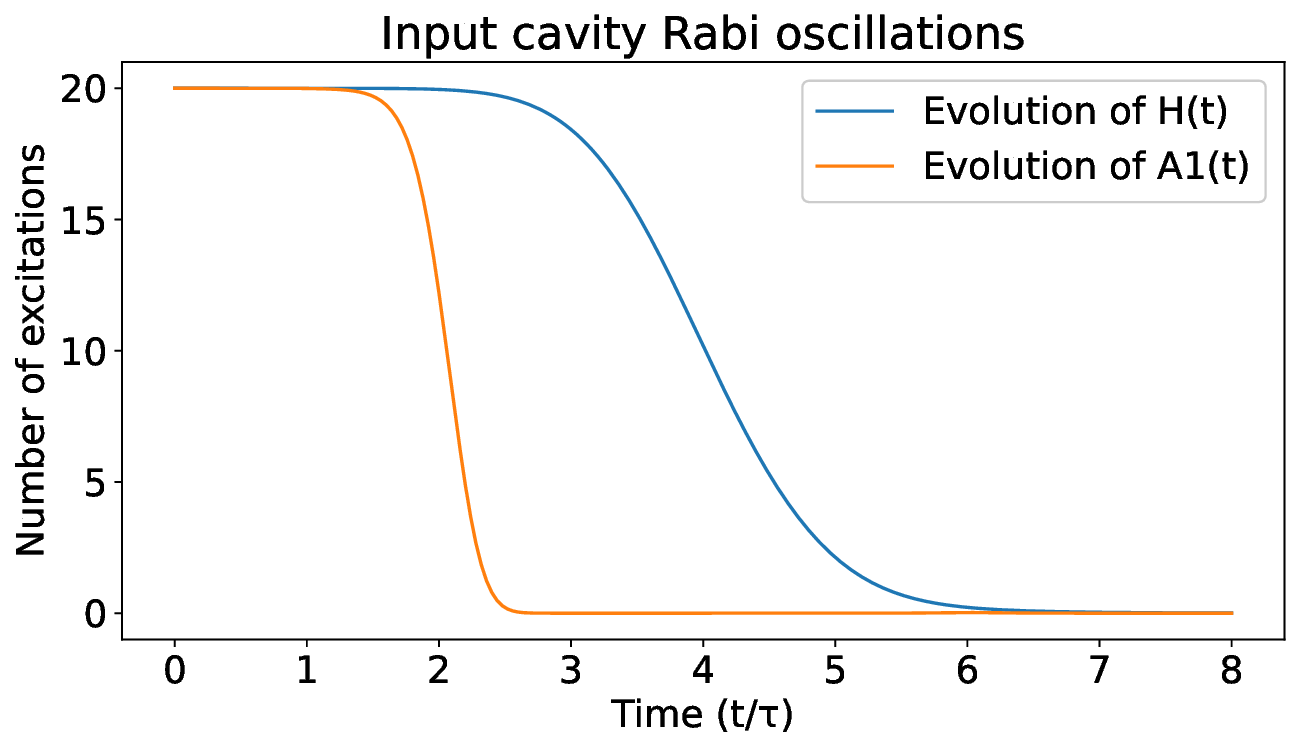} \\
    \includegraphics[width=1\columnwidth]{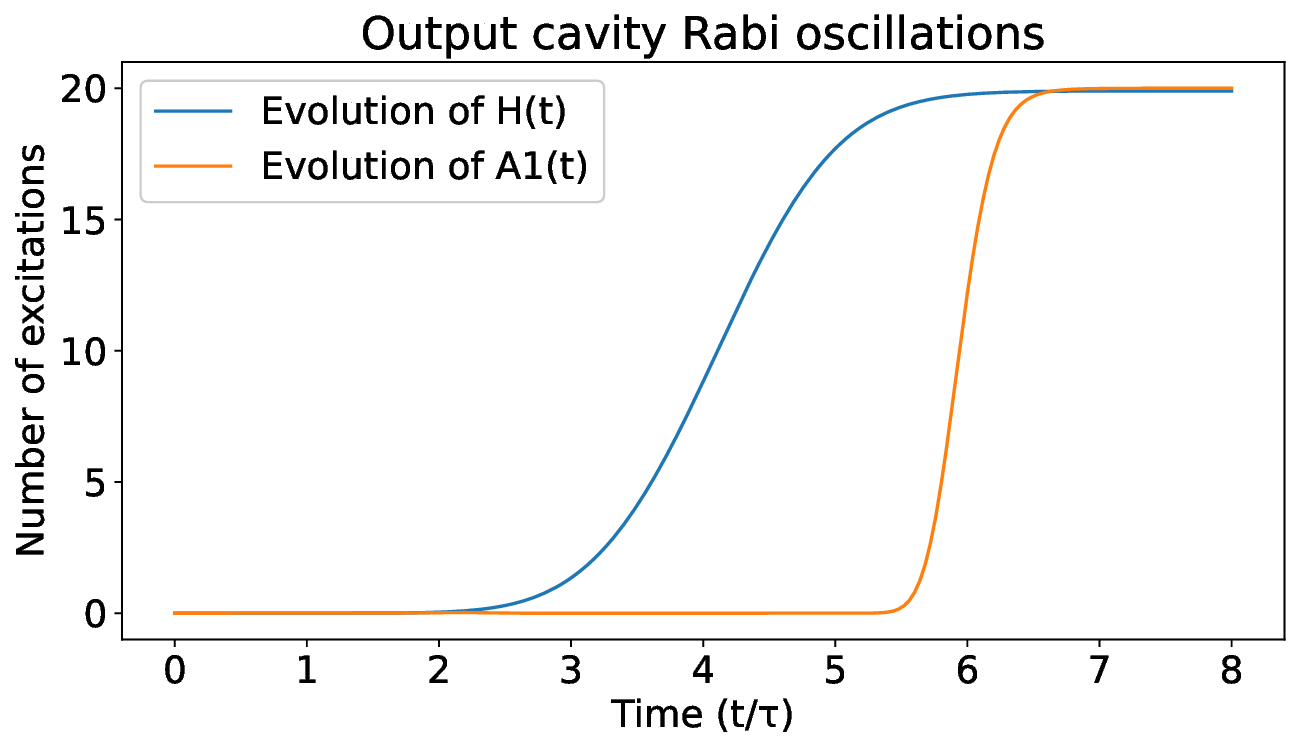}
\end{tabular}
    \caption{Population dynamics for $n=20$ and $\gamma=0.15/\tau$. Top: atomic population. Middle: input-cavity population. Bottom: output-cavity population.}
    \label{fig-adiabat_non_adiabat_populations_20}
\end{figure}
\begin{figure}[tbph]
    \includegraphics[width=1\columnwidth]
    {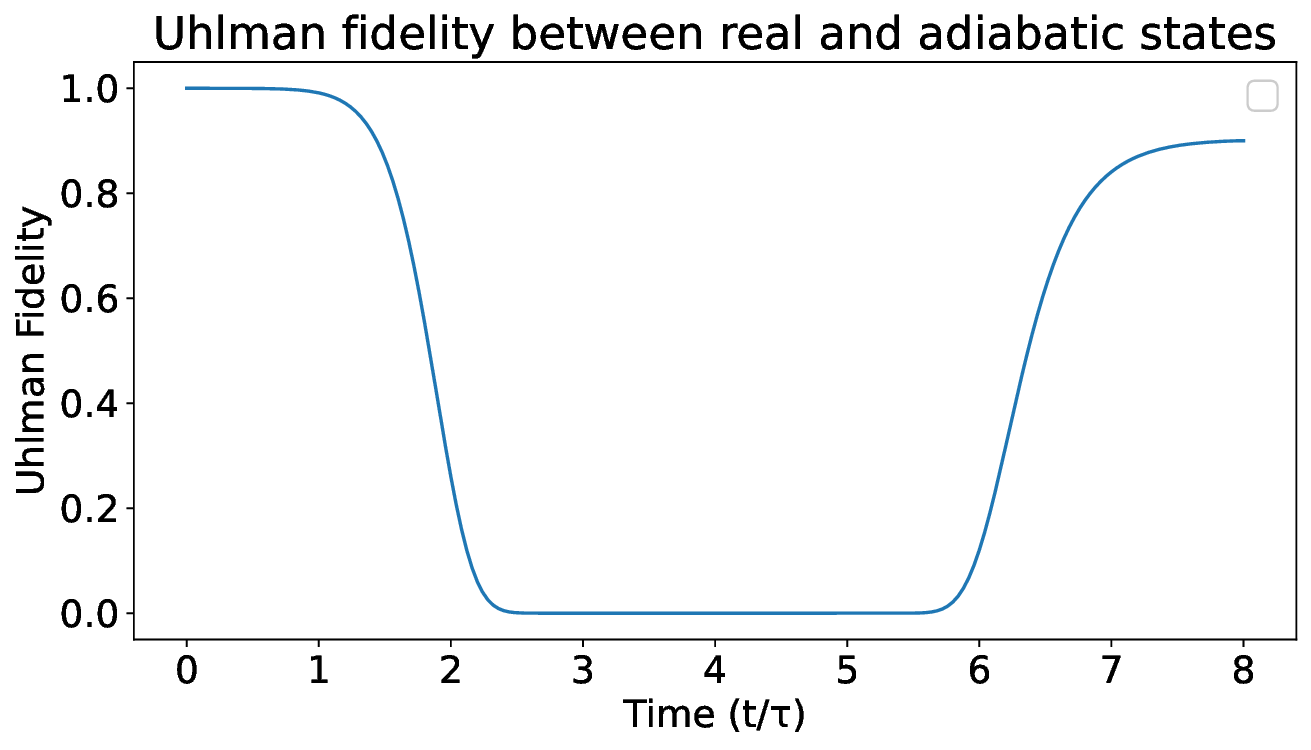}
    \caption{Fidelity of states for $n=20$ and $\gamma=0.15/\tau$.}
    \label{fig-adiabat_non_adiabat_fidelity_20}
\end{figure}

Figures~\ref{fig-adiabat_non_adiabat_populations_20} and \ref{fig-adiabat_non_adiabat_fidelity_20} show the same population and fidelity comparisons, respectively, for $n = 20$ and $\gamma = 0.15/\tau$ (note that the product of $n$ and $\gamma$ is preserved). The interval of disagreement is broader, as suggested by Fig.~\ref{fig-adiabatic-bounds}(bottom), but the final overlap with the reference is higher.

 The traveling-pulse Hamiltonian therefore has a genuine connection with STIRAP, but the oscillatory dynamics shown here is not complete adiabatic dark-state transfer. The three-state diagnostic and normal-mode comparisons illustrate the coupling structure, with the limitations stated above. At fixed $n\gamma$, increasing $n$ makes following the exact many-photon dark state more demanding, even though the initial atomic drive matrix element remains fixed.

\section{Conclusions \label{Sec:conclusions}}
We have demonstrated that coherent and Fock states of a traveling pulse can produce qualitatively different atomic dynamics. In the Gaussian-pulse examples studied here, coherent input produces damped Rabi oscillations, whereas Fock input exhibits a partial recovery of the oscillation amplitude in the trailing part of the pulse. Extending the comparison discussed in Ref.~\cite{molmer_2024}, our analytical derivations and QuTiP simulations reveal a transient response that depends on the quantum state of the incident radiation. This recovery is distinct from the conventional collapse-and-revival dynamics of a coherent field in a lossless cavity.

The transformed master equation exposes the role of the new virtual output cavity (buffer) and the competing atom--field coupling terms. Their different weights provide a physical interpretation of how excitation stored during the interaction can contribute to the late-pulse recovery. The comparison with the Hamiltonian-only dynamics further shows that dissipation substantially affects the recovery amplitude. These results establish the importance of treating coherent exchange and radiative loss together when interpreting the atomic response to a quantized traveling pulse.

We have also established a connection with STIRAP at the Hamiltonian level through an exact many-photon dark state. Its structure and adiabaticity condition clarify the limits of the three-state analogy: the observed oscillatory dynamics occurs away from ideal dark-state following, and Hamiltonian darkness alone does not ensure protection against radiative loss. At fixed $n\gamma$, following the many-photon dark state becomes more demanding as the photon number increases. The distinction between transient recovery and adiabatic transfer provides a basis for investigating pulse shapes and coupling regimes that enhance the recovery or approach controlled transfer between temporal modes.

\section*{Acknowledgments}
We dedicate this work to the memory of Joseph H. Eberly, whose pioneering work has shaped our understanding of coherent atom-field dynamics.
This research is supported by the Bulgarian NSF grant ``Quantum-Control Engineered Selectivity for Sensing, Localisation and Robust Gates''. 

\nocite{*}
\bibliography{references} 

\end{document}